\documentclass[11pt,a4paper]{article}
\usepackage{MySty}
\usepackage{amsmath,amsfonts,amssymb,graphics}

\usepackage[normalem]{ulem}
\usepackage{url}
\usepackage{graphicx}
\usepackage{cancel}
\usepackage{mathrsfs}
\usepackage{bbm}
\usepackage{pifont}
\usepackage{braket}
\usepackage{siunitx}
\usepackage{slashed}
\usepackage{mathtools}
\usepackage{xspace}
\usepackage{booktabs}
\usepackage{csquotes}
\usepackage[dvipsnames,svgnames]{xcolor}
\usepackage{tikz}
\usetikzlibrary{positioning,shapes,arrows}
\usetikzlibrary{decorations.pathmorphing} 
\usetikzlibrary{decorations.markings}
\usepackage{standalone}
\usepackage{blkarray}
\usepackage{url}
\usepackage{orcidlink}

\usepackage{tikz-feynman}
\tikzfeynmanset{compat=1.1.0}
\usetikzlibrary{arrows}
\tikzset{
    vertex/.style={circle,draw, minimum size=1.5em},
    edge/.style={->,> = latex'}
}
\tikzset{
  small dot/.style={dot, minimum size=3pt, inner sep=0pt}, % You can adjust size here
}

\usepackage[normalem]{ulem}

\def\r {\rightarrow}

\newcommand{\bmt}{\begin{pmatrix}}
\newcommand{\emt}{\end{pmatrix}}
\newcommand{\ba}{\begin{array}{c}}
\newcommand{\ea}{\end{array}}
\newcommand{\be}{\begin{equation}}
\newcommand{\ee}{\end{equation}}
\newcommand{\bea}{\begin{eqnarray}}
\newcommand{\eea}{\end{eqnarray}}

\newcommand{\bi}{\begin{itemize}}
\newcommand{\ei}{\end{itemize}}

\newcommand{\baz}{\begin{array}{cc}}
\newcommand{\mathsym}[1]{{}}

\newcommand{\bt}{\begin{tabular}}
\newcommand{\et}{\end{tabular}}

\newcommand{\benu}{\begin{enumerate}}
\newcommand{\eenu}{\end{enumerate}}

\newcommand{\matindex}[1]{\mbox{\scriptsize#1}}

\begin{document}

\title{Universal Seesaw Leptogenesis}

\author[a]{K.S. Babu\,\orcidlink{0000-0001-6147-5155},}
\affiliation[a]{Department of Physics, Oklahoma State University, Stillwater, OK 74078, USA}
\emailAdd{babu@okstate.edu}

\author[b]{Maximilian Berbig\,\orcidlink{0000-0002-8585-6099},}
\affiliation[b]{Instituto de Física Corpuscular (CSIC-Universitat de València),
Parc Científic UV, C/Catedrático José Beltrán, 2, E-46980 Paterna, Spain}
\affiliation[b]{Departament de Física Teòrica, Universitat de València, 46100 Burjassot, Spain}
\emailAdd{berbig@ific.uv.es}

\author[c]{Srubabati Goswami\,\orcidlink{0000-0002-5614-4092},}
\affiliation[c]{Theoretical Physics Division, Physical Research Laboratory, Ahmedabad-380009, India}
\emailAdd{sruba@prl.res.in}

\author[d]{and Drona Vatsyayan\,\orcidlink{0000-0002-6868-3237}}
\affiliation[d]{Department of Physics, Carleton University, Ottawa, ON K1S 5B6, Canada}
\affiliation[d]{Arthur B. McDonald Canadian Astroparticle Physics Research Institute, 64 Bader Lane, Queen's University, Kingston, ON K7L 3N6, Canada}
\emailAdd{drona@physics.carleton.ca}

\date{\today}
\abstract{
We study the implications for leptogenesis in a class of left-right symmetric model, where all fermion masses are induced through the Universal Seesaw mechanism. Unlike conventional analyses, we do not use the decays of the neutrino embedded in the right-chiral lepton doublet, but rather those of the gauge-singlet mediators required for neutrino mass generation in the canonical Type-I seesaw. This model features a  generalized parity symmetry, which is motivated by the solution to the strong $CP$ problem. Since this discrete symmetry doubles the fermionic degrees of freedom in this model, we can generate the required $CP$ violation in the heavy fermion decays with only a single generation of mediators. One of the distinct features of our scenario is that the bounds from thermalization or washout via gauge interactions typically encountered in the canonical  left-right symmetric models do not apply here. Moreover, the heavy mediators can decay to both the left and the right-chiral neutrinos, leading to a cancellation in the resulting baryon asymmetry for decays above the left-right symmetry breaking scale. We discuss ways to avoid this cancellation and show that low scale left-right symmetry breaking above the current collider limits  $v_R >\SI{18}{\tera\electronvolt}$ is viable. The right chiral neutrinos also obtain their masses from the seesaw mechanism, and the lightest one turns out to have a sub-eV scale mass. We find that its abundance is consistent with standard cosmology, and it acts as potentially observable dark radiation.}

\maketitle

%%%%%%%%%%%%%%%%%%%%%%%%%%%%
\section{Introduction}\label{sec:intro}
Left-right symmetric models based on the gauge group  $\text{SU}(3)_C\otimes \text{SU}(2)_L \otimes \text{SU(2)}_R  \otimes \text{U}(1)_{X}$ \cite{Pati:1974yy,Mohapatra:1974hk,Mohapatra:1974gc,Senjanovic:1975rk,Mohapatra:1980yp} provide one of the most straightforward gauge extensions of the Standard Model (SM). Left-right symmetry (LRS) can also serve as an intermediate stage symmetry on the road to grand unification in SO(10) \cite{Fritzsch:1974nn,Georgi:1974my}. Here, $\text{U}(1)_{X}$ can in general be distinct \cite{Bolton:2019bou} from the commonly used $\text{U}(1)_{B-L}$, which gauges the difference between baryon number ($B$) and lepton number $(L)$.

While the left-right symmetric model with all SM fermions in doublets of the respective gauge groups is inherently chiral, one can define a discrete exchange symmetry that transforms each bosonic and fermionic representation of one weak gauge sector into its conjugate of the opposing sector. Such a \textit{generalized Parity}  symmetry $\mathcal{P}$ bears the advantage of forbidding any $CP$ violating topological operator such as the QCD $\theta$-term. In fact, if one breaks the LRS and the electroweak symmetry only via the Vacuum Expectation Values (VEV) of  scalar doublets $H_R$ and $H_L$ for each SU(2) factor, one can engineer a situation in which the phases of the quark mass matrices of each sector cancel, thus solving the strong $CP$ problem  \cite{Babu:1989rb} (see also Refs.~\cite{Hall:2018let,Hall:2019qwx,Craig:2020bnv,deVries:2021pzl,Benabou:2025viy} for more recent investigations). The price one has to pay is the introduction of vector-like pairs of SU(2) singlet fermions (or bidoublet fermions \cite{Hall:2018let}) to generate the masses of all quarks and leptons via the Universal Seesaw mechanism \cite{Berezhiani:1983hm,Chang:1986bp,Davidson:1987mh,Rajpoot:1986nv,Babu:1988mw,Babu:1989rb}. Since the resulting fermion masses are quadratic in the Yukawas, this approach bears the advantage that one can explain all quark and charged lepton masses with couplings ranging from  $\mathcal{O}(10^{-3}) - \mathcal{O}(1)$, in contrast with the range $\mathcal{O}(10^{-6}) -\mathcal{O}(1)$ needed in the SM. For  applications to the flavor problem, see Refs.~\cite{Babu:2018vrl,Dcruz:2022rjg,Dcruz:2023mvf,Jana:2021tlx,Jana:2024icm}. 

However, since the Higgs doublet VEVs $\braket{H_L^0} = v_L$ and $\braket{H_R^0}= v_R$ do not break $\text{U}(1)_{B-L}$, this setup does not predict the nature of the active neutrino masses, hence, one can accommodate both Majorana and 
(pseudo-)Dirac Neutrinos. Parametrically small Dirac neutrino masses can even be realized without the inclusion of heavy fermionic messengers in the neutrino sector via loop diagrams \cite{Mohapatra:1987nx,Babu:1988yq,Balakrishna:1988bn,Borah:2017leo,Babu:2022ikf,Babu:2023dzz}. 

Moreover, the Universal Seesaw framework offers rich applications for cosmology, such as providing dark matter candidates \cite{Dunsky:2019api,Dunsky:2019upk,Dror:2020jzy,Dunsky:2020dhn,Baldwin:2025oqt}, implementations of leptogenesis \cite{Gu:2010yf,Dunsky:2020dhn,Babu:2024glr,Carrasco-Martinez:2023nit,Borboruah:2025bwx} and potentially detectable Gravitational waves from the underlying phase transition \cite{Dunsky:2019upk,Craig:2020bnv,Graf:2021xku,Dasgupta:2025uzi}.

In this work, we focus on a novel realization of leptogenesis \cite{FUKUGITA198645} in such models. Previous studies relied on the  decays of the right-handed neutrino (RHN) $\nu_R$ housed in right-chiral lepton doublet $l_R$, that obtains its large mass from either the VEV $v_R$ of the $\text{SU}(2)_R \otimes \text{U}(1)_X$ breaking doublet in the Universal Seesaw
 \cite{Gu:2010yf,Dunsky:2020dhn,Carrasco-Martinez:2023nit} or from the VEV of an $\text{SU}(2)_R$ triplet (see e.g. Refs.~\cite{Joshipura:2001ya,Rodejohann:2002mh}) via the Type II Seesaw mechanism \cite{Lazarides:1980nt,Schechter:1980gr,Mohapatra:1980yp,Cheng:1980qt,Wetterich:1981bx}
in the  conventional left-right symmetric model \cite{Gunion:1989in,Deshpande:1990ip,Zhang:2007fn,Zhang:2007da}.
We instead choose to investigate the decays of the additional super-heavy messenger fields $N_{L,R}$ that are integrated out to yield the Type I Seesaw \cite{Minkowski:1977sc,Yanagida:1979as,Gell-Mann:1979vob,Glashow:1979nm,Yanagida:1980xy, PhysRevLett.44.912}   Majorana masses for both the active neutrinos $\nu_L$, and $\nu_R$. Since $N_{L,R}$ are gauge singlets, and we consider $\nu_R$ to be a sub-eV state instead, we automatically avoid constraints from gauge mediated equilibration and washout that usually enforce high scale LRS breaking   \cite{Ma:1998sq,Frere:2008ct,BhupalDev:2014hro,Gu:2017gra}.

Unlike the previous mentioned scenarios, our approach can take place either \textit{before} or after LRS breaking, which allows us to accommodate both high scale and low scale LRS. We find that the required size of $v_R$ for the high-scale regime is in tension with limits from the quality of the $\mathcal{P}$-based solution to the strong $CP$ problem, which further motivates the phenomenologically more accessible low scale regime.  

Due to the  parity symmetry,  there will be two distinct fermions $N_L$ and $N_R$ per generation, and we show that even a single generation of them is enough to source the prerequisite $CP$ violation via the interference of tree level and one loop vertex and self-energy corrections. However, the downside of our construction is that we need to break $\mathcal{P}$ \textit{softly} at the level of $N_{L,R}$ masses because $CP$-violation requires them to be non-degenerate. 

Furthermore, since each messenger field can decay to both  lepton doublets $l_L$ and $l_R$, we observe that a cancellation can arise for decays at temperatures above the LRS breaking scale $v_R$: For temperatures below $\mathcal{O}(\SI{e11}{\giga\electronvolt})$, the sphaleron transitions of both $\text{SU}(2)_L$ and $\text{SU}(2)_R$ are in thermal equilibrium. As $l_{L,R}$ carry opposite charges under $\text{U}(1)_{B-L}$, the resulting chiral baryon asymmetries thus sum to zero. We can avoid this outcome by considering a \textit{hard} breaking of $\mathcal{P}$ in the neutral lepton Yukawas, allowing us to either generate different amounts of $CP$ violation in the decays to $l_L$ and $l_R$, or selectively washout the asymmetry stored in $l_R$. Such a hard breaking of $\mathcal{P}$ in the lepton sector does not directly upset the parity solution to the strong $CP$ problem, since no such hard breaking is assumed in the quark sector.

Our analysis indicates that successful leptogenesis is possible for $W_R$ gauge boson masses $m_{W_R}$ as low as the current collider limit of 6 TeV \cite{ATLAS:2019erb}, and we further delineate the cosmologically allowed parameter space of the lightest RHN $\nu_R$ with a sub-eV mass.

The manuscript is organized as follows. We introduce the Universal Seesaw model in Section \ref{sec:model} and review the ingredients to the $\mathcal{P}$-based solution to the strong $CP$ problem in Section \ref{sec:strongCP}. Section \ref{sec:numass} deals with neutrino mass generation. After introducing the generalities for leptogenesis in Section \ref{sec:LeptoGeneral}, we discuss the
cases of decays before and after LRS breaking together with all possible sources of washouts in Section \ref{sec:scenarios}. The cosmology of the lightest RHN is the focus of Section \ref{sec:RHN}. We provide a summary of our results and discuss their implications in Section \ref{sec:concl}. All the 
relevant  decay widths and branching ratios can be found in Appendix \ref{app:BRs}, and our chemical potential analysis for the spectator processes in the presence of the two $\text{SU}(2)_{L,R}$ sphalerons can be found in Appendix \ref{app:chem}.
Appendix \ref{sec:WObreak} is dedicated to an alternative mechanism for washing out the right chiral lepton abundance. 

%%%%%%%%%%%%%%%%%%%%%%%%%%%%%
\section{The Model}\label{sec:model}
\subsection{Overview}

\begin{table}[t!]
    \centering
    \begin{tabular}{|c||c|c|c|c||c|c|}
        \hline
          Fields & $\text{SU}(3)_c$ & $\text{SU}(2)_L$ & $\text{SU}(2)_R$ & $B-L$ & $\zeta$ & $X = B-L + \zeta$\\
          \hline
          \hline 
          $q_L$ & \textbf{3} & \textbf{2} & \textbf{1} & ${1}/{3}$ & 0 &${1}/{3}$\\
          $l_L$ & \textbf{1} & \textbf{2} & \textbf{1} & $-1$ & 0 &  $-1$\\
          $H_L$ & \textbf{1} & \textbf{2} & \textbf{1} & 0&1 & 1\\
          \hline 
          \hline 
          $q_{R}^\dagger$ & \textbf{3}& \textbf{1} & \textbf{2}  & ${1}/{3}$& 0 &${1}/{3}$ \\
          $l_R^\dagger$ & \textbf{1} & \textbf{1} & \textbf{2}  & $-1$ & 0 & $-1$\\
          $H_R^\dagger$& \textbf{1} & \textbf{1} & \textbf{2} & 0& 1 & $1$ \\
          \hline
          \hline
          $U_{L},\; U_R^\dagger$ & \textbf{3} & \textbf{1} & \textbf{1} & ${1}/{3}$ & 1 & ${4}/{3}$ \\
          $D_{L},\; D_R^\dagger$ & \textbf{3} & \textbf{1} & \textbf{1} & ${1}/{3}$ & $-1$ & $-{2}/{3}$ \\
            $E_{L},\;E_R^\dagger$ & \textbf{1} & \textbf{1}& \textbf{1}& $-1$ & $-1$& $-2$ \\
          \hline
          $N_{L},\;N_R^\dagger$ & \textbf{1}& \textbf{1}& \textbf{1}& $-1$ & 1 & 0\\
          \hline
    \end{tabular}
    \caption{Gauge charges of all fields in the Universal Seesaw model. The \textit{first (second)} block contains the $\text{SU}(2)_L\;(\text{SU}(2)_R)$ doublets and the \textit{third} block contains pairs of vector-like fermions that are all singlets under  $\text{SU}(2)_L\otimes \text{SU}(2)_R$. All fields without (with) a $\dagger$ are left-chiral (right-chiral) Weyl spinors and we utilize the two component spinor notation of Ref.~\cite{Dreiner:2008tw}.}
    \label{tab:fields}
\end{table}

We consider a model based on the gauge group $\text{SU}(3)_C \otimes \text{SU}(2)_L \otimes \text{SU}(2)_R \otimes \text{U}(1)_{X}$, whose generators lead to the following relation for the electric charge $Q$:
\begin{equation}
    Q = T_{L_3} + T_{R_3} + \frac{X}{2}= T_{L_3} + Y\,,
\end{equation} 
where $Y$ denotes the SM hypercharge.

 The particle content of the Universal Seesaw model \cite{Berezhiani:1983hm,Chang:1986bp,Davidson:1987mh,Rajpoot:1986nv,Babu:1988mw,Babu:1989rb}  contains the usual standard model fermions along with RHNs and four sets of vector like fermions ($U_i, D_i, E_i, N_i$), where $i$ denotes the fermion generation. In the usual LR symmetric model one can identify $\text{U}(1)_{X}$ with $\text{U}(1)_{B-L}$, since $\text{SU}(2)_R \otimes \text{U}(1)_{X}$ is broken by the VEV of a $\text{SU}(2)_R$ triplet \cite{Gunion:1989in,Deshpande:1990ip}, sourcing the mass of $\nu_R$ embedded in $l_R$ via the Type II Seesaw mechanism \cite{Lazarides:1980nt,Schechter:1980gr,Mohapatra:1980yp,Cheng:1980qt,Wetterich:1981bx} and hence breaks $B-L$ by two units. In the case of Universal Seesaw, there is no scalar whose VEV breaks $B-L$ and the only source of breaking will come from hard Majorana mass terms for the gauge singlet fermions $N_{L,R}$ and lepton number-violating couplings to the doublet fields. Additionally, $\text{U}(1)_{X}$  acts as hypercharge $\text{U}(1)_{Y}$  on the vector-like singlets ($U_i, D_i, E_i, N_i$) and consequently the $X$ charge is distinct from the $B-L$ charges of these fields. Therefore, we follow the convention of Ref.~\cite{Bolton:2019bou}, and decompose the $X$ charge as follows
\begin{align}
    X= B-L + \zeta\,,
\end{align}
in terms of $B-L$ and an auxiliary charge $\zeta$. The representations and charges of all matter fields can be found in Table~\ref{tab:fields}. We further assume a generalized global parity symmetry under which all fields transform as 
\begin{align}
    \psi_L(t,\vec{x}) \leftrightarrow \psi_R^\dagger (t,-\vec{x})\,,
\end{align}
which can be understood as a generalized notion 
of space-time parity $\mathcal{P}$. Here $\psi_{L,R}$ is a generic shorthand notation for all species of fermions, the Higgs doublets $H_{L,R}$ as well as the $W_{L,R}$ gauge bosons.

We consider the  following Yukawa interactions  
\begin{align}
-\mathcal{L}_\text{Yukawa} =  &\left(Y_u\right)_{ij} q_{L_i} \varepsilon H_L U_{R_j} + \left(Y_d\right)_{ij} q_{L_i} H_L^\dagger D_{R_j} + \left(Y_e\right)_{ij} l_{L_i} H_L^\dagger E_{R_j}\nonumber \\
  + &\left(Y_u'\right)_{ij} q_{R_i} \varepsilon H_R U_{L_j} + \left(Y_d'\right)_{ij} q_{R_i} H_R^\dagger D_{L_j} + \left(Y_e'\right)_{ij} l_{R_i} H_R^\dagger E_{L_j} + \text{H.c.}\,,
\end{align}
where $\varepsilon\equiv i \sigma_2$ is defined in terms of the second Pauli matrix and the vector-like mass terms for the messenger fields are given by
\begin{align}\label{eq:MU}
   \left(M_U\right)_{ij} U_{L_i} U_{R_j} + \left(M_D\right)_{ij} D_{L_i} D_{R_j} + \left(M_E\right)_{ij} E_{L_i} E_{R_j} + \text{H.c.}\,.
\end{align}
Generalized parity leads to the following relations:
\begin{align}
    \left(Y_f\right)_{ij} = \left(Y_f'\right)^{*}_{ij}, \quad \left(M_F\right)_{ij}=\left(M_F\right)^{\dagger}_{ij} \quad \text{with} \quad  f= u,d,e,\quad F=U,D,E\,.
\end{align}
The charged fermion masses are generated from the Universal Seesaw mechanism after the neutral components of both Higgs doublets $H_{L,R}$ condense at the scales $v_L$ and $v_R$ respectively \cite{Berezhiani:1983hm,Chang:1986bp,Davidson:1987mh,Rajpoot:1986nv,Babu:1988mw,Babu:1989rb} 

\begin{align}
    m_f  \simeq \frac{v_L v_R}{2} Y_f^{*} M_F^{-1} Y_f^T  \quad \text{with} \quad  f= u,d,e\quad \text{and~} F = U,D,E\,.
\end{align}
One can assign, for example, a common mass $M$ to all the messenger fields and generate the observed spectrum of fermion masses by adjusting $Y_f$'s. The perturbativity of the bottom quark Yukawa couplings imply that the scale of the vector-like masses satisfies \cite{Craig:2020bnv}
\begin{align}\label{eq:pert}
    \frac{v_R}{M} \gtrsim \mathcal{O}(1\%)\,.
\end{align}
On the other hand, the mass of the vector-like quark that mediates the top-quark mass can not be larger than $v_R$, because one needs to reproduce the top's order one Yukawa coupling to the SM like Higgs boson. At leading order in the Seesaw expansion: $v_{L,R}/M_F$, the fermion mass eigenstates arrange themselves into light fields $f$ and heavy fields $F$ 
\begin{align}
    \hat{f} = \begin{pmatrix}
        f_L\\
        f_R^\dagger
    \end{pmatrix},
  \quad \hat{F} = \begin{pmatrix}
      F_L\\
      F_R^\dagger 
  \end{pmatrix}\,,
\end{align}
where $f=u,d,e$ and $F=U,D,E$. The above holds for the charged leptons and lighter quarks.

\subsection{Mechanisms for Parity Breaking}
In this section we introduce the known mechanism for generating a hierarchical spectrum of vevs $v_R\gg v_L$, without choosing a particular realization.

The part of the Higgs potential, which respects a global $\mathcal{O}(8)$ custodial symmetry \cite{Chacko:2005pe}, of which only its subgroup $\text{SU}(4)$ will be relevant for us, can be written in terms of  $\mu^2>0,\; \lambda>0$ as 
\begin{align}
    V_{LR} \supset -\mu^2 (|H_L|^2 +  |H_R|^2) + \frac{\lambda}{2}(|H_L|^2 + |H_R|^2)^2\,.
\end{align}
First, one can break the custodial symmetry by adding the term  
\begin{align}
    V_{\slashed{\text{SU}(4)}} = \lambda' |H_L|^2 |H_R|^2\,,
\end{align}
with $\lambda'>0$, that leads to the solutions 
\begin{align}
v_L=0 \land v_R\neq 0,\; v_L \neq 0 \land v_R=0,\; v_L=v_R\neq 0\,.    
\end{align}
The hierarchy $v_L=\SI{246}{\giga\electronvolt}\ll v_R$ dictated by phenomenology requires additional custodial symmetry breaking contributions, see Ref.~\cite{Jung:2019fsp} for a systematic discussion.  In fact, one can understand the Higgs boson aligned with the smaller VEV $v_L$ as the pseudo-Nambu-Goldstone boson of the explicitly broken custodial $\text{SU}(4)$ \cite{Chacko:2005pe}. 

Without adding new fields, the aforementioned hierarchy can be realized by adding a soft generalized parity breaking mass $\mu_\text{soft}^2>0$ for $H_R$ that also breaks the custodial symmetry \cite{Babu:1988mw,Babu:1989rb}
  \begin{align}\label{eq:soft}
    -\mu_\text{soft}^2 |H_R|^2\,, 
\end{align}
which results in 
\begin{align}
    v_L = \sqrt{\frac{\mu^2}{2\lambda+\lambda'} - \frac{\lambda+\lambda'}{\lambda'(2\lambda+\lambda')}\,\mu_\text{soft}^2}, \quad 
    v_R = \sqrt{\frac{\mu^2}{2\lambda+\lambda'} + \frac{\lambda}{\lambda'(2\lambda+\lambda')}\,\mu_\text{soft}^2}\,.
\end{align}
A small $v_L$ can then arise  by tuning the two contributions $\propto \mu^2$ and $\propto \mu_\text{soft}^2$ against each other.

An alternative without introducing a source of soft breaking consists of fine-tuning $\lambda'$ against radiative corrections dominated by the top sector \cite{Hall:2018let,Hall:2019qwx,Carrasco-Martinez:2025zus}, which also break custodial symmetry.
The advantage of this approach is that it predicts $v_R$ in terms of the scale $\mathcal{O}(\SI{e11}{\giga\electronvolt})$ at which the  quartic coupling of the SM like Higgs vanishes. Furthermore, in the absence of any soft-breaking there exists only a single dimensionful term in the scalar potential, like in the SM,  and it can be shown that this implies the same amount of fine-tuning as in the SM \cite{Hall:2018let}. 

Another approach to parity breaking introduces a parity odd, real singlet scalar $S$ \cite{Babu:1989rb} that spontaneously breaks the $\mathcal{Z}_2$ by developing a VEV  of $v_S=\mu_S/\sqrt{2\lambda_S}$ \cite{Chang:1983fu,Chang:1984uy}
\begin{align}\label{eq:PotS}
    V_S &= - \mu_S^2 S^2 + \frac{\lambda_S}{4} S^4 
     + \lambda_{HS} S^2 (H_L^\dagger H_L + H_R^\dagger H_R)  - \kappa S (H_R^\dagger H_R - H_L^\dagger H_L)\nonumber \,.
\end{align}
After $S$ condenses, the term proportional to $\kappa>0$ sources a mass splitting
\begin{align}
    \mu_{L\;\text{eff.}}^2 = \mu^2 - \kappa v_S + \lambda_{HS} v_S^2\,,\quad
    \mu_{R\;\text{eff.}}^2 = \mu^2 + \kappa v_S + \lambda_{HS} v_S^2\,,
\end{align}
from which 
\begin{align}
    v_L= \sqrt{\frac{\mu^2}{2\lambda + \lambda'} - \frac{\kappa v_S}{\lambda'} },\quad 
    v_R = \sqrt{\frac{\mu^2}{2\lambda + \lambda'} + \frac{\kappa v_S}{\lambda'} }
\end{align}
follows, and it is evident that now $v_L \ll v_R$ can be realized by tuning the contribution $\propto \mu^2$ against the one $\propto \kappa v_S$. 
We identify the lighter boson $h_L$ mostly contained in the neutral component of $H_L$ as the SM Higgs boson with a  mass of 125 GeV. We denote the heavier Higgs as $h_R$ and  depending on the other potential parameters, its mass can be close to $v_R$. Note that the presence of a parity odd singlet can jeopardize the solution to the strong CP problem at the loop level or via higher dimensional operators, as discussed in Sections \ref{sec:parity} and \ref{sec:qual}.

In the following we focus on the cosmological implications of the explicit and spontaneous symmetry breaking scenarios.
 
\subsection{Collider Limits}
Due to the generalized parity, the ratios of the massive gauge bosons are fixed to be
\begin{align}
    \frac{m_{Z'}}{m_Z} \simeq  \frac{m_{W_R}}{m_W} = \frac{v_R}{v_L}\,. 
\end{align}
Dilepton searches at the LHC rule out $m_{Z'} <\SI{5}{\tera\electronvolt}$ \cite{ATLAS:2019erb}, which translates to a lower bound of $v_R >\SI{13}{\tera\electronvolt}$ \cite{Craig:2020bnv}. Similarly, direct searches for a single charged lepton and missing transverse momentum from $pp \rightarrow W_R^\pm \rightarrow l_R^\pm \overset{(-)}{\nu_R}$ can probe the production of $W_R^\pm$, and  exclude $m_{W_R} < \SI{6}{\tera\electronvolt}$ \cite{ATLAS:2019lsy}, leading to the stronger bound~\cite{Craig:2020bnv}\footnote{Dijet searches carried out by the CMS collaboration are slightly less constraining as they only rule out $m_{W_R} < \SI{3.6}{\tera\electronvolt}$ \cite{CMS:2019gwf}. }
\begin{align}\label{eq:18}
    v_R >\SI{18}{\tera\electronvolt}\,.
\end{align}
It should be noted that the bounds from searches for (same or opposite sign) dileptons together with dijets \cite{ATLAS:2018dcj,CMS:2018agk}, that apply to the usual LR symmetric model with a heavy $\nu_R$ decaying into the second charged lepton, do not apply here, as we consider $\nu_R$'s that are effectively massless from the point of view of collider physics.

Note that the deviations of the Higgs couplings to top quark compared to SM are negligible.  This follows from the diagonalization of the seesaw quark mass matrix. The effective theory reduces to the SM, with deviations of the couplings to the fermion $f$ arising as higher dimensional operators of the schematic form $f_L H_L^\dagger H_L f_R/\Lambda$ in terms of $\Lambda=\text{Max}[v_R,M_F]$, where $M_F$ is the mass of the corresponding vector-like fermion. However these corrections are negligible as long as  $\text{Max}[v_R,M_F] \gtrsim \mathcal{O}({\rm TeV})$: In Ref.~\cite{Dcruz:2023mvf} constraints on the Higgs couplings from top decays, Higgs decays to muons or taus as well as meson mixing were analyzed. The large Yukawa couplings we assume are well within the existing bounds for  vector-like messenger fields at or above the TeV-scale.

\section{Strong CP problem}\label{sec:strongCP}
\subsection{Solution from Generalized Parity}\label{sec:parity}
The only $CP$-violating term in the QCD Lagrangian is given by 
\begin{align}
    \left(\underbrace{\theta + \text{Arg}(\text{det}[ \mathcal{M}_u \mathcal{M}_d]}_{\equiv\, \overline{\theta} })\right) \tilde{G}_{\mu\nu} G^{\mu \nu},
\end{align}
where $\tilde{G}_{\mu \nu} = \epsilon_{\mu \nu \alpha \beta} G^{\alpha \beta}$ with the Levi-Cevita pseudo-tensor $\epsilon_{\mu \nu \alpha \beta}$, and $\mathcal{M}_{u,d}$ denote the up- and down-type quark mass matrices.
Parity removes the $CP$-odd term $\theta \tilde{G}_{\mu\nu} G^{\mu \nu}$, so that only the phases of the quark mass matrices need to be considered. The freedom to perform $\text{SU}(2)_L \times \text{SU}(2)_R \times \text{U}(1)_X$ gauge transformations (see, for example, Ref.~\cite{Zhang:2007da}) is used to render the VEVs of $H_{L,R}$ real \cite{Babu:1989rb}. The  $6\times 6$ quark mass matrices $\mathcal{M}_{u,d}$ have the following structure 
\begin{align}
         \mathcal{M}_u = \begin{blockarray}{ccc}
    \matindex{$u_R$} & \matindex{$U_R$} & \\
    \begin{block}{(cc)c}
      \mathbb{O}_{3\times 3} &  \frac{v_L}{\sqrt{2}} Y_u & \matindex{$u_L$} \\
      \frac{v_R}{\sqrt{2}} Y_u^\dagger& M_U & \matindex{$U_L$} \\
    \end{block}
  \end{blockarray}\,,
\end{align}
where we focus on the up-quark sector for brevity, and the mass matrices of the vector-like quarks were defined in Eq.~\eqref{eq:MU}. The upper left zero block arises from the gauge structure and because we only consider renormalizable couplings. Owing to the generalized parity $\mathcal{P}$, one finds that 
\begin{align}
    \text{Arg}(\text{det}[ \mathcal{M}_u]) &= \frac{v_L v_R}{2} \left(\text{Arg}(\text{det}[Y_u])  + \text{Arg}(\text{det}[Y_u^\dagger]) \right) =0\,,
\end{align}
and the same result holds for the down-type quarks. 
In Ref.~\cite{Babu:1989rb}, it was shown that one-loop contributions to $\text{Arg}(\text{det}[ \mathcal{M}_{u,d}])$ vanish. The authors of Ref.~\cite{deVries:2021pzl} demonstrated that corrections can occur at two-loops if the generalized parity is softly broken: this can occur either when the tree-level mass matrix of the vector-like fermions is non-Hermitian \;$M_{U,D}\neq M_{U,D}^\dagger$ or from the phenomenologically unavoidable mass splitting in the scalar potential $\mu_R^2\neq \mu_L^2$. 
The contribution from the vector-like fermions is by far the larger one \cite{deVries:2021pzl}.
However, note that while the soft scalar mass term is almost necessary for a viable phenomenology, there is no reason to consider non-Hermitian fermion masses, and these fermionic terms will never be induced by the soft breaking in the scalar sector. 

For the case of spontaneously broken parity with the parity odd singlet scalar $S$, there exists an additional $CP$ violating contribution to the vector-like quark masses from the allowed operators
\begin{align}
    i S \left(y_U U_L U_R + y_D D_L D_R \right) + \text{H.c.}\,,
\end{align}
after $S$ condenses. At tree level these terms do not contribute to $\overline{\theta}$ due to the upper left zero block in $\mathcal{M}_{u,d}$ in Eq.~\eqref{eq:MU}. However, there occurs a shift of $\overline{\theta}$ at one loop of the order \cite{Craig:2020bnv}
\begin{align}
    \delta \overline{\theta} \simeq \frac{v_S}{16\pi^2 M}(|y_U|+ |y_D|)\,,
\end{align}
which together with the limit from the perturbativity of the bottom quark Yukawa couplings in Eq.~\eqref{eq:pert}, and assuming  $v_S\simeq v_R$ for simplicity, implies that \cite{Craig:2020bnv}
\begin{align}\label{eq:smallYuk}
    |y_U|+ |y_D| \lesssim \mathcal{O}(10^{-6})\,. 
\end{align}

\subsection{Quality Problem}\label{sec:qual}

The solution to the strong $CP$ problem could be jeopardized by the presence of higher dimensional operators that correct the entries of the tree-level quark mass matrices. This caveat was first discussed in Ref.~\cite{Berezhiani:1992pq} and recently re-evaluated in Ref.~\cite{Craig:2020bnv}. 

If the generalized parity symmetry we employ is a global symmetry, it is expected to be explicitly violated by quantum gravity effects such as wormholes or virtual black holes \cite{Georgi:1981pu,Dine:1986bg,Coleman:1989zu,Abbott:1989jw,Holman:1992us,Kamionkowski:1992mf,Barr:1992qq,Ghigna:1992iv,Kallosh:1995hi,Alonso:2017avz}. One finds that the lowest parity-\textit{breaking} contribution arises at dimension 5 \cite{Berezhiani:1992pq}
\begin{align}\label{eq:danger}
    \frac{1}{M_\text{Pl.}} q_L\left(\alpha_u \varepsilon H_L H_R \varepsilon + \alpha_d H_L^\dagger H_R^\dagger \right)q_R\,,
\end{align}
where parity is explicitly broken for $\alpha_i \neq \alpha_i^\dagger,\; (i=u,d)$, and the largest contribution comes from the correction to the lightest quark masses.
The strong $CP$ problem is still solved as long as \cite{Craig:2020bnv}
\begin{align}
    v_R < \frac{\SI{20}{\tera\electronvolt}}{|\alpha|} \left(\frac{\overline{\theta}}{10^{-10}}\right),
\end{align}
which in the light of Eq.~\eqref{eq:18} would only leave a small parameter space. 

The situation changes once we assume that the generalized parity is the residual symmetry of a gauge symmetry. Since gauge symmetries are the only exact symmetries in a quantum gravity context, explicit symmetry breaking from Planck-suppressed higher dimensional operators is absent. Nevertheless higher dimensional operators that shift $\overline{\theta}$ can still arise even without parity breaking:
At dimension six one finds \cite{Hall:2018let}
\begin{align}
    \frac{\beta}{M_\text{Pl.}^2} \left(|H_L|^2-|H_R|^2\right) G_{\mu \nu}\tilde{G}^{\mu\nu},
\end{align}
where $G$ denotes the QCD field strength and the limit on $v_R$ reads \cite{Hall:2018let}
\begin{align}\label{eq:paritybound}
    v_R < \frac{\SI{e13}{\giga\electronvolt}}{\sqrt{|\beta|}} \sqrt{\frac{\overline{\theta}}{10^{-10}}}\,.
\end{align}
Note that if this model is embedded in $\text{SO}(10)\otimes \textit{CP}$ one can compute $\beta$ and it is found that $\beta\ll 1$ \cite{Hall:2018let}. 
If $\mathcal{P}$ is broken spontaneously by the VEV of a parity-odd scalar $S$ then the following dimension five terms will break $CP$ \cite{DAgnolo:2015uqq}
\begin{align}\label{eq:SSB1}
        \frac{i S}{M_\text{Pl.}} \left( \gamma_u q_L H_L U_R + \gamma_u^* q_R H_R U_L +  
        \gamma_d q_L H_L^\dagger D_R   + \gamma_d^* q_R H_R^\dagger D_L\right) + \text{H.c.}\,,
\end{align}
and one requires \cite{Craig:2020bnv}
\begin{align}\label{eq:qual2}
    v_S\simeq v_R < \frac{\SI{e7}{\giga\electronvolt}}{|\gamma|}\left(\frac{\overline{\theta}}{10^{-10}}\right).
\end{align}
Throughout this work we  either assume that the operator in Eq.~\eqref{eq:danger} is sufficiently suppressed to lift $v_R$ above $\mathcal{O}(\SI{10}{\tera\electronvolt})$ or that generalized parity is a discrete gauge symmetry in order to avoid the bound in Eq.~\eqref{eq:paritybound}, which can be motivated in theories with extra space-time dimensions \cite{Choi:1992xp}, such as string theory \cite{Dine:1992ya}. It is important to mention that while there exist examples of gauged $CP$, the same is not necessarily true for generalized parity. 

\subsection{Domain Walls}\label{sec:DW}
It is well known  that the spontaneous breakdown of the discrete generalized parity can give rise to the formation of approximately two dimensional topological defects known as domain walls \cite{Zeldovich:1974uw}. Domain walls from LRS breaking were studied in Refs.~\cite{Cline:2002ia,Mishra:2009mk,Banerjee:2020zxw,Borboruah:2022eex}. 
Since these macroscopic field configurations have an energy density that redshifts slower than both radiation and matter, they eventually dominate the energy budget of the universe. An early phase of domain wall dominated expansion proceeds as power-law inflation and is incompatible with BBN and CMB decoupling \cite{Gelmini:1988sf}, because radiation and matter would have been diluted too much. Even if the energy density of the wall never dominates in the early universe, there still exist limits from the gravitational lensing of CMB photons \cite{Zeldovich:1974uw, Friedland:2002qs}, which should not lead to temperature fluctuations larger than the primordial modes observed in the CMB. This domain wall problem lead to the common assumption that LRS is broken before the end of inflation so that the walls are diluted by the exponential expansion during inflation. 

An alternative consists of considering $\mathcal{P}$ to be only approximate so that the walls can disappear dynamically. First let us focus on the minimal model with $H_{L,R}$. Here the defect forms in the neutral components $h_{L,R}$ of the Higgs doublets $H_{L,R}$, because there  exist two degenerate minima in $\sqrt{h_L^2+h_R^2}$  and these minima correspond to the choice of which doublet picks up the large VEV $v_R$ \cite{Jung:2019fsp}. As argued in Ref.~\cite{Rai:1992xw,Carrasco-Martinez:2023nit}, a small explicit breaking of the generalized parity can make these walls collapse by generating a pressure that drives them towards each other \cite{Kibble:1976sj,Vilenkin:1981zs,Sikivie:1982qv}. Further, if the explicit breaking is large enough so that it leads to an energy difference between the formerly degenerate vacua larger than the barrier separating them, no walls are formed in the first place \cite{Riva:2010jm}. Thus, we take the scenario with the generalized parity being softly broken by the term in Eq.~\eqref{eq:soft} to be safe from the domain wall problem. 

If global $\mathcal{P}$ is instead spontaneously broken by the condensate of the parity odd $S$ in Eq.~\eqref{eq:PotS} the defect is formed in the field $S$ instead. Again we can add an additional bias term $S^n+\text{H.c.}\;(n
>0)$ to destabilize the walls \cite{Rai:1992xw} or prevent their formation to begin with.

On the other hand, if we assume that $\mathcal{P}$ is a gauge symmetry, no explicit breaking is allowed because in quantum gravity gauge symmetries are exact \cite{Krauss:1988zc}. As argued in Refs.~\cite{McNamara:2022lrw,Asadi:2022vys}, a dynamical mechanism for the disappearance of a defect in space-time symmetry such as $CP$ - or in our case $\mathcal{P}$ - would need to dynamically change the topology of space-time, which is not expected to happen even in a theory of quantum gravity. Thus, for the case of gauged discrete parity we can only appeal to inflation to get rid of the walls, which roughly implies $T_\text{RH}<v_R$ \cite{McNamara:2022lrw,Asadi:2022vys}, where $T_\text{RH}$ is the post-inflationary reheating temperature.

%%%%%%%%%%%%%%%%%%%%%%%%%%%%%
\section{Neutrino Masses}\label{sec:numass}
Considering the neutrino sector, the relevant Yukawa interactions read 
\begin{align}\label{eq:41}
   -\mathcal{L}_\text{Yukawa}^\nu =  Y_{LR} l_L \varepsilon H_L N_R 
    + Y_{RL} l_R \varepsilon H_R N_L 
    + Y_{LL} l_L \varepsilon H_L N_L + Y_{RR} l_R \varepsilon H_R N_R + \text{H.c.}\,,
\end{align}
and the hard mass terms of the sterile fields $N_{L,R}$ are given by 
\begin{align}
    M_{LR} N_L N_R 
    +\frac{1}{2} M_{LL} N_L N_L + \frac{1}{2} M_{RR} N_R  N_R + \text{H.c.}\,,
\end{align}
where $M_{LR}$ denotes the Dirac mass term and $M_{LL,RR}$ denote the Majorana mass terms. The terms involving $M_{LL,RR}$ and $Y_{LL,RR}$ break $B-L$ by two units.
Note that  the generalized parity symmetry  implies 
\begin{align}
    &\left(Y_{LR}\right)_{ij} = \left(Y_{RL}\right)_{ij}^*,\quad  \left(Y_{LL}\right)_{ij} = \left(Y_{RR}\right)_{ij}^*,\nonumber\\
    &\left(M_{LL}\right)_{ij} =\left(M_{RR}\right)_{ij}^\dagger, \quad \left(M_{LR}\right)_{ij} = \left(M_{LR}\right)_{ij}^\dagger.
\end{align}
However we take the above relations only to be approximate, which will be motivated in sections \ref{sec:CP} and \ref{sec:lowscale}.

We take $l_L\;(l_R)$ to be the lepton doublet aligned with the lightest active (sterile) neutrino mass eigenstate and introduce only a single generation of both $N_L$ and $N_R$. 
In the following we focus on the lightest generations and denote them as  
\begin{align}
    \nu_{L,R} \equiv \nu_{L,R}^{(1)}, \quad N_{L,R}\equiv N_{l,R}^{(1)}.
\end{align}
In this single generation picture, the {$4\times 4$} mass matrix  in the basis $(\nu_L, \nu_R, N_L, N_R)$ is given by 
\begin{align}\label{eq:matnu}
   M_\nu = \begin{pmatrix}
       0 & 0 & Y_{LL} v_L & Y_{LR} v_L \\
       0 & 0 &Y_{LR}^* v_R & Y_{LL}^* v_R \\
       Y_{LL}  v_L & Y_{LR}^* v_R & M_{LL} & M_{LR} \\
       Y_{LR} v_L & Y_{LL}^* v_R & M_{LR} & M_{RR}
   \end{pmatrix} 
\end{align}
and here all Yukawa couplings are just complex numbers.
The Yukawa couplings can be parameterized as 
\begin{align}
    Y_{LR} \equiv y \sin(\alpha) e^{-\frac{i \varphi}{2}},\quad
    Y_{LL} \equiv y \cos(\alpha) e^{\frac{i \varphi}{2}},
\end{align}
where $\varphi$ is a CP phase and $\alpha$ takes into account that the magnitude of the modulus of each coupling can differ. In order to reduce the number of free parameters we take $\alpha=\pi/4$ and define 
\begin{align}\label{eq:Y}
    Y\equiv  y \sin\left(\frac{\pi}{4}\right).
\end{align}
Here, we take $M_{LL,RR}$ to be real and without loss of generality we can always find a basis with  $M_{LR}=0$. We  integrate out $N_{L,R}$, whose masses read
\begin{align}
    m_{N_L} \simeq M_{LL}\,,\quad 
    m_{N_R} \simeq M_{RR}\,,
\end{align}
and then diagonalize the resulting $2\times2$ matrix in the space of $(\nu_L,\nu_R)$, which leads to  the following absolute values (in the limit $M_{LL,RR}\gg v_R \gg v_L$)
\begin{align}
    m_{\nu_L} &\simeq  \frac{3 Y^2 v_L^2}{8 M_{LL}} \sqrt{1+\frac{1}{r}+\frac{2\cos{(2\varphi)}}{\sqrt{r}}}\,,\label{eq:mnuL}\\
    m_{\nu_R} &\simeq \frac{Y^2 (v_L^2+4 v_R^2)}{8 M_{LL}} \sqrt{1+\frac{1}{r}+\frac{2\cos{(2\varphi)}}{\sqrt{r}}}\label{eq:mnuR}\,,
\end{align}
where we defined
\begin{align}
    r\equiv \frac{M_{RR}^2}{M_{LL}^2}\,.
\end{align}
The ratio of the masses therefore read as
\begin{align}\label{eq:ratio}
    \frac{m_{\nu_L}}{m_{\nu_R}} \simeq \frac{3 v_L^2}{4 v_R^2}\,,
\end{align}
and the mixing angle between $\nu_L$ and $\nu_R$ is found to be
\begin{align}\label{eq:mix}
    \theta_{LR} \simeq -\frac{v_L}{2 v_R}\,.
\end{align}
The relation for the $\nu_R$ mass in Eq.~\eqref{eq:ratio} is important for its cosmological implications that will be analyzed in Section \ref{sec:RHN}. 

Since we neglect all couplings to the heavier generations of $N_{L,R}$ and $l_{L,R}$, we assume that the observed mass splittings of active neutrinos {$\nu_L^{(2.3)}$} are sourced by the heavier two generations {$N_{L,R}^{(2,3)}$}.\footnote{{When it comes to lepton mixing we take the following perspective: The PMNS angles can be accounted for entirely by the charged lepton sector. In the neutral lepton sector we assume (approximately) decoupled $4\times4$ matrices of the schematic form of Eq.~\eqref{eq:matnu}, so the heavier $N_{L,R}^{(2,3)}$ do not impact the mass of $\nu_{L,R}^{(1)}$ and the masses of  $\nu_{L,R}^{(2,3)}$ are independent of $N_{L,R}^{(1)}$.\label{footnote}}}

Consequently, we demand that the lightest neutrino is effectively massless
\begin{align}\label{eq:light}
    m_{\nu_L} < \sqrt{\Delta m_{21}^2}= \SI{8.7e-3}{\electronvolt}\,,
\end{align}
where we use $\Delta m_{21}^2= \SI{7.49e-5}{\electronvolt\squared}$ \cite{Esteban:2024eli}.

\section{Leptogenesis}\label{sec:LeptoGeneral}

\begin{figure}[!t]
    \centering
    \begin{tikzpicture}
    \begin{feynman}
    \vertex(i1) {\color{blue}{$N_i$}};
    \vertex[right=1.25 cm of i1,small dot] (i2){};
    \vertex[right=1.25 cm of i2] (e);
    \vertex[above=0.5 cm of e] (a) {$l_{k}$};
    \vertex[below=0.5 cm of e](b) {$H_k$};
    \diagram* {
    (i1)--[color=blue] (i2)-- (a),
    (i2)-- [scalar](b),
    };
    \end{feynman}
    \end{tikzpicture}
    \hspace{2em}
    \begin{tikzpicture}
    \begin{feynman}
    \vertex(i1) {\color{blue}{$N_i$}};
    \vertex[right=1.25 cm of i1,small dot] (i2){};
    \vertex[right=1.25 cm of i2] (e);
    \vertex[above=0.7 cm of e,small dot] (a){};
    \vertex[below=0.7 cm of e,small dot](b){};
    \vertex[right=1.0 cm of a](c) {$l_{k}$};
    \vertex[right=1.0 cm of b](d) {$H_k$};
    \diagram* {
    (i1)--[color=blue] (i2)--[scalar, edge label = $H_k$] (a),
    (i2)-- [edge label'=\(l_{k}\)](b),
    (a)--[edge label =\(N_j\),color=red] (b),
    (a)-- (c),
    (b)-- [scalar](d)
    };
    \end{feynman}
    \end{tikzpicture}
    \hspace{2em}
    \begin{tikzpicture}
    \begin{feynman}
    \vertex(i1) {\color{blue}{$N_i$}};
    \vertex[right=1.0 cm of i1,small dot] (i2){};
    \vertex[right=1.0 cm of i2,small dot] (f){}; 
    \vertex[right=0.75 cm of f,small dot] (g){};
    \vertex[right=1.0 cm of g] (e);
    \vertex[above=0.5 cm of e] (a) {$l_{k}$};
    \vertex[below=0.5 cm of e](b) {$H_k$};
    \diagram* {
    (i1)--[color=blue] (i2)--[half left,edge label=\(l_{m}\)] (f) --[scalar, half left, edge label = $H_m$] (i2),
    (f)--[edge label' =\(N_j\),color=red] (g),
    (a)-- (g)-- [scalar] (b),
    };
    \end{feynman}
    \end{tikzpicture}
    \caption{Tree-level \textit{(left)}, one-loop vertex \textit{(middle)} and self-energy \textit{(right)} diagrams for the decays of heavy fermions $N_i \r l_k H_k$ with $\{i,j,k,m\} = L,R$ and $i \neq j$.}
     \label{fig:lepto}
\end{figure}
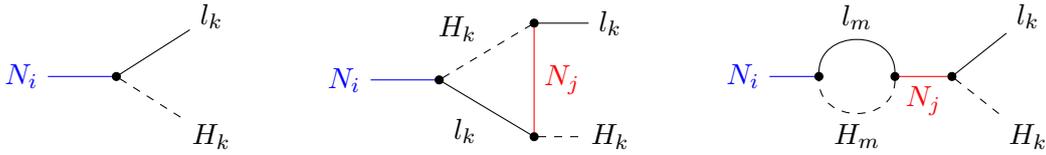

It is useful to define the  dimensionless abundances $Y_X$ and asymmetries $\Delta_X$ of the particle species $X$ as
\begin{align}
    Y_X = \frac{n_X}{s}, \quad  Y_X + Y_{\overline{X}} \simeq 2  Y_X, \quad \Delta_X =Y_X - Y_{\overline{X}}\,,
\end{align}
where the asymmetries are related to the chemical potentials $\mu_X$ via 
\begin{align}
    \frac{\Delta_X}{Y_X^\text{eq.}} = {\frac{2\pi^2}{9 \zeta(3)}} \frac{\mu_X}{T}{\simeq 2\frac{\mu_X}{T}}\,.
\end{align}
Throughout this work we use the superscript \enquote{eq.} to denote quantities evaluated in thermal equilibrium. 

\subsection{CP Violation}\label{sec:CP}
The CP violation comes from the interference of the tree and the loop-level diagrams for the decays of $N_{L,R}$, as shown in Fig.~\ref{fig:lepto}.\footnote{{Here we do not consider loops of $N_{L,R}^{(2,3)}$, since we work in a regime where they have negligible coupling to $\nu_{L,R}$ as indicated in footnote \ref{footnote}. For the same reason we only consider decays of $N_{L,R}$ to $l_{L,R}$ instead of to $l_{L,R}^{(2,3)}$. Therefore the Davidson-Ibarra bound \cite{Davidson:2002qv} does not apply here.}} It can be seen that we have contributions from diagrams with massive states in both the loops and the final states as well as diagrams with massive states either in the loop or the final states. Note that in the vertex correction diagram, the states $l_k$ and $H_k$ in the loop should match the final states, whereas, for the self energy diagram, they can be same or different. 

For concreteness, we will only consider the mass of the right-handed Higgs $H_R$, which will lead to phase space suppression in some decays and for notational convenience we define 
\begin{align}
    \delta_{L,R} \equiv \frac{m_{h_R}^2}{M_{LL,RR}^2}\,, 
\end{align}
such that $\delta_R = {\delta_L}/{r}$.

Throughout this work, we take $r>1$.
The loop integrals for the sum of self-energy and vertex corrections in the case of two masses ($H_R$ is present in both the loop and the final state) read \cite{Hugle:2018qbw}
\begin{align}
    f^{(2)}(r,\delta) =\sqrt{r}\left(\frac{(1-\delta)^2}{1-r}+ 1 + \frac{1-2\delta +r }{(1-\delta)^2}\text{Log}\left(\frac{r-\delta^2}{1-2\delta +r}\right)\right)\,,
\end{align} 
and for a single massive state ($H_R$ in the loop and $H_L$ in the final state) there is only a self-energy diagram 
\begin{align}
    f^{(1)}(r,\delta) = \sqrt{r}\frac{1-\delta}{1-r}\,.
\end{align}
Obviously $f^{(2)}$ reduces to the standard result in the limit of $\delta \rightarrow 0$ 
\begin{align}
    f(r) = \sqrt{r}\left(\frac{1}{1-r}+ 1+ (1+r)\text{Log}\left(\frac{r}{1+r}\right)\right)\,.
\end{align}
Here $\delta$ can be $\delta_L$ or $\delta_R$. In order to have a well-defined $CP$ asymmetry, we need to softly break the exchange symmetry $M_{LL}\neq M_{RR}$. This is not expected to affect the solution to the strong $CP$ problem via generalized parity, as the heavy Majorana neutrinos only couple to the quark sector at higher loop order. We typically assume that $M_{LL}$ and $M_{RR}$ have the same order of magnitude but do not consider them to be almost degenerate. We will not consider the regime of  quasi degenerate masses for resonant leptogenesis and take $1.1 \leq r \leq 25$ for our analysis. All decay widths and branching ratios can be found in Appendix \ref{app:BRs}.

The $CP$-violating decay parameters for decays to $N_{L,R}\rightarrow l_L H_L $  are found to be
\begin{align}\label{eq:cpasym1}
   \varepsilon^L_L \equiv    \varepsilon_L(N_L) &=
   \frac{\text{Im}\left(\left(Y_{LL}Y_{LR}^*\right)^2\right)}{8\pi (|Y_{LR}|^2 (1-\delta_L)^2+|Y_{LL}|^2)} \left(f\left(r\right)  + f^{(1)}\left(r,\delta_L\right)\right)\,,\nonumber\\
    \varepsilon^L_R  \equiv  \varepsilon_L(N_R) &=
    -\frac{\text{Im}\left(\left(Y_{LL}Y_{LR}^*\right)^2\right)}{8\pi (|Y_{LR}|^2+|Y_{LL}|^2 (1-\delta_R)^2)} \left(  f\left(\frac{1}{r}\right) + f^{(1)}\left(\frac{1}{r},\delta_R\right)\right)\,,
\end{align}
and for decays to $N_{L,R}\rightarrow  l_R H_R $ one obtains instead 
\begin{align}\label{eq:cpasym2}
       \varepsilon^R_L \equiv \varepsilon_R(N_L) &=\frac{\text{Im}\left(\left(Y_{LL}Y_{LR}^*\right)^2\right)}{8\pi (|Y_{LR}|^2 (1-\delta_L)^2+|Y_{LL}|^2)} \left(f^{(2)}\left(r,\delta_L\right)+ f^{(1)}\left(r,\delta_L\right)\right)\,,\nonumber\\
       \varepsilon^R_R \equiv  \varepsilon_R(N_R) &=-\frac{\text{Im}\left(\left(Y_{LL}Y_{LR}^*\right)^2\right)}{8\pi (|Y_{LR}|^2+|Y_{LL}|^2 (1-\delta_R)^2)} \left(
       f^{(2)}\left(\frac{1}{r},\delta_R\right) + f^{(1)}\left(\frac{1}{r},\delta_R\right) \right)\,.
\end{align}
It is evident that the total $CP$ violation in each channel vanishes for $M_{LL}=M_{RR}\;(r=1)$,\footnote{To see this, note that $\delta_L=\delta_R$ for $M_{LL}=M_{RR}$ and due to Eq.~\eqref{eq:Y} we always have $|Y_{LR}|^2=|Y_{LL}^2|=Y^2$.}
\begin{align}\label{eq:Weinberg}
 \varepsilon_L^L+  \varepsilon_R^L=0 =   \varepsilon_L^R +   \varepsilon_R^R\,,
\end{align}
in accordance with the findings of Ref.~\cite{Nanopoulos:1979gx}. The sign difference between $\varepsilon^L_L$ and $\varepsilon^L_R$ (or $\varepsilon^R_L$ and $\varepsilon^R_R$) arises due to the couplings being complex conjugated. 
We checked that there is an irreducible phase  
\begin{align}
    \varphi  \equiv \text{Arg}\left(Y_{LL} Y_{LR}^*\right)\,,
\end{align}
by exhausting all possible field redefinitions. 
The presence of both $N_L$ and $N_R$ due to parity doubling is what allows for successful leptogenesis in this model, where we only incorporate a single generation of leptons. The parity partner plays the role of the second generation of Majorana neutrinos that is required for $CP$ violating decays. 

\subsection{Out-of-Equilibrium Dynamics}
The Hubble rate $H(T)$ and the entropy density of the plasma $s(T)$ during radiation domination read 
\begin{align}
    H(T) = \sqrt{\frac{8\pi^3 g_\rho(T)}{90}} \frac{T^2}{M_\text{Pl.}}, \quad s(T)= \frac{2\pi^2}{45} g_S(T) T^3\,,
\end{align}
where $g_\rho\;(g_S)$ denotes the number of relativistic degrees of freedom in energy (entropy). We utilize the fitting functions for the temperature dependence of $g_{\rho,S}$ of Ref.~\cite{Wantz:2009it}. 
\subsubsection{Overview}

We assume that the two heavier generations of $N_{L,R}$ are too heavy to be produced and therefore we assume the following hierarchy of scales
\begin{align}
    M_{{LL,RR}_{2,3}} > T_\text{RH} > M_{LL,RR}\,,
\end{align}
where $T_\text{RH}$ is the post-inflationary reheating temperature of the radiation dominated thermal bath.
The limit on the inflationary Hubble rate $H_I<\SI{6e13}{\giga\electronvolt}$ from the non-observation of inflationary tensor modes~\cite{Planck:2018jri}, allows to  constrain  the reheating temperature $T_\text{RH}\sim \sqrt{H_I M_\text{Pl.}}$ and consequently 
\begin{align}
    M_{LL,RR} < \SI{6.5e15}{\giga\electronvolt}\,.
\end{align}
Throughout this work we assume a vanishing initial abundance for both $N_{L,R}$ and in the spirit of thermal leptogenesis their populations are produced entirely via their Yukawa couplings.

The out-of-equilibrium dynamics are encoded in the efficiency factors $\kappa_{L,R}$, and are determined by numerically solving the Boltzmann equations, discussed below.
However, we first present some analytical estimates for the efficiency factors following e.g. Refs.~\cite{Buchmuller:2004nz,Davidson:2008bu}. In general, it is convenient to define the decay parameter for each $N_{L,R}$ from the decay widths summarized in Appendix \ref{app:BRs}
\begin{align}
    K_{L} &\equiv \frac{\Gamma(N_{L} \rightarrow l_L H_L, l_L^\dagger H_L^\dagger)  + \Gamma(N_{L} \rightarrow l_R H_R, l_R^\dagger H_R^\dagger)}{H(T=M_{LL})},\nonumber\\
    K_R &\equiv \frac{\Gamma(N_{R} \rightarrow l_L H_L, l_L^\dagger H_L^\dagger)  + \Gamma(N_{R} \rightarrow l_R H_R, l_R^\dagger H_R^\dagger)}{H(T=M_{RR})} = \frac{K_L}{\sqrt{r}}\,.
\end{align}
In the following we will use the notation $M_{LL}=M_N$ and for the moment we take $r\simeq 1$. 
When $K_{L,R} \ll 1$, which corresponds to 
\begin{equation}\label{eq:ooE1}
   Y \ll 6 \times 10^{-3} \sqrt{\frac{M_N}{\SI{e12}{\giga\electronvolt}}}\,, 
\end{equation}
for $\delta_{L,R}\ll1$ and $g_{\rho}(M)= \mathcal{O}(100)$, we are in the so-called weak washout regime. Here the Yukawa couplings are too small for $N_{L,R}$ to develop a thermal abundance and the asymmetry is produced from out-of-equilibrium decays at $T\ll M_N$. This regime is sensitive to the initial abundance of $N_{L,R}$, and for a vanishing initial abundance one finds that the number density of the produced $N_{L,R}$ is proportional to $K_{L,R}$ \cite{Buchmuller:2004nz}. The number of decaying $N_{L,R}$ is also proportional to $K_{L,R}$, which is why here one finds $\kappa_{L,R}\simeq K_{L,R}^2$. 

On the other hand, we are in the strong washout regime for $K_{L,R} \geq 1$, where the Yukawa couplings are large enough to generate a thermal population of $N_{L,R}$. Since thermal equilibrium is an attractor, this regime is insensitive to the  initial abundances of $N_{L,R}$.  Here, the leptonic asymmetry remains constant once  the washout from inverse decays eventually decouple at $T\simeq M_N/10$. In this case, one can show up to the logarithmic accuracy $\kappa_{L,R}\simeq 1/K_{L,R}$, see for example, Ref.~\cite{Giudice:2003jh,Davidson:2008bu}. 

To summarize, one can parameterize the order of magnitude of the efficiency as 
\begin{align}
    \kappa_{L,R} \simeq 
    \begin{cases}
    K_{L,R}^2 \quad &\text{for}\quad K_{L,R}\ll1\\
    \frac{1}{K_{L,R}} \quad &\text{for}\quad K_{L,R}\geq1
    \end{cases}\,.
\end{align}

\subsubsection{Boltzmann Equations}\label{sec:BE}
We work with the following set of coupled Boltzmann Equations (BEqs) for the abundance of $N_{L,R}$ and the leptonic asymmetries $\Delta_{L,R}$
\begin{align}\label{eq:beq}
    sHz\frac{dY_{N_{i}}}{dz} &= -\gamma_{N_{i}}\left(\frac{Y_{N_{i}}}{Y_{N_{i}}^{\rm eq.}}-1\right)\,,\nonumber\\
    sHz\frac{d\Delta_L}{dz} &= \sum_{i=L,R}\gamma_{N_{i}}\left[\varepsilon_{i}^L\left(\frac{Y_{N_{i}}}{Y_{N_{i}}^{\rm eq.}}-1\right)-{\rm Br}_i^L \frac{\Delta_L}{2Y_{l_L}^{\rm eq.}} \right]+(2 \leftrightarrow 2\; \text{washout/transfer})\,,\nonumber\\
    sHz\frac{d\Delta_R}{dz} &= \sum_{i=L,R}\gamma_{N_{i}}\left[\varepsilon_{i}^R\left(\frac{Y_{N_{i}}}{Y_{N_{i}}^{\rm eq.}}-1\right)-{\rm Br}_i^R\frac{\Delta_R}{2Y_{l_R}^{\rm eq.}}\right]+(2 \leftrightarrow 2\; \text{washout/transfer})\,,
\end{align}
where $z \equiv M_{N}/T$. The total decay rate is given by
\begin{align}
    \gamma_{N_i}= n_{N_i} \frac{K_1(z)}{K_2(z)}\Gamma_{N_i}\,,
\end{align}
where the total widths are 
\begin{equation}
    \Gamma_{N_i} = \sum_{j=L,R}\Gamma(N_i \r l_j H_j) + \Gamma(N_i \r l_j^\dagger H_j^\dagger)\,,
\end{equation}
and the CP asymmetries $\varepsilon_i^{j}~(i,j=L,R)$ are defined in Eqs.~\eqref{eq:cpasym1} and \eqref{eq:cpasym2}.

The $2 \leftrightarrow 2$ washout/transfer correspond to the off-shell part of the $2\leftrightarrow 2$ scatterings 
such as $l_L H_L \leftrightarrow l^\dagger_L H^\dagger_L$, $l_R H_R \leftrightarrow l^\dagger_R H^\dagger_R$ ($\Delta L =2$, washout) and $l_L H_L \leftrightarrow l^\dagger_R H^\dagger_R$ ($\Delta L =0$, transfer), which we discuss below. We perturbatively expand in the small couplings $Y_{LR},\; Y_{LL}$ necessitated by the out-of-equilibrium conditions. Therefore, our analysis focuses on the decays and inverse decays of $N_{L,R}$, which are Boltzmann suppressed at low temperatures, and we ignore all scattering processes involving  $N_L N_L,\; N_R N_R,\; N_L N_R$ distributed across initial and final states  because their couplings are suppressed by the fourth power of the relevant Yukawas, and they are double Boltzmann suppressed at low temperatures. 

As discussed above, our scenario is similar to thermal leptogenesis with hierarchical RHNs. Hence, for the regime we are interested in, the Yukawa interactions quickly bring $N_{L,R}$ in thermal equilibrium at $z<1$ due to the production from inverse decays. The inverse decays go out of equilibrium for $z \gtrsim 1$, and the decays of $N_{L,R}$ source the lepton asymmetries. The lepton asymmetries eventually freeze-in as the number densities of $N_{L,R}$ become Boltzmann suppressed.  

We solve the BEqs in Eq.~\eqref{eq:beq} numerically to obtain the values of $\Delta_i$ at $z \gg 1$, taking vanishing initial abundance for both $N_L$ and $N_R$, as well as for the asymmetries $\Delta_{L,R}$. 

The total baryon asymmetry depends on the regime we are in, and is given by\footnote{Note that $l_L$ and $l_R$ have opposite charges under $B-L$ of $Q_{(B-L)_L}=-Q_{(B-L)_R}=-1$, see table \ref{tab:fields}.}
\begin{equation}
    \Delta_B = c_\text{sph.} \left(Q_{(B-L)_L} \Delta_{L} +Q_{(B-L)_R} \Delta_{R} \right) =\frac{28}{79}\begin{cases}
        \left(\Delta_{L} - \Delta_{R}\right)\quad &\text{for}\quad T>v_R\\
       \Delta_{L} \quad &\text{for}\quad T<v_R
    \end{cases}\,,
\end{equation}
 where the two distinct scenarios corresponding to $T > v_R$ and $T < v_R$ are discussed in detail in Section~\ref{sec:scenarios}.

\subsection{Washout Processes}

Now, we discuss the conditions under which the washout from various processes, that were not accounted for in the Boltzmann equations in Eq.~\eqref{eq:beq} can be neglected.

\subsubsection{Yukawa Interactions}

Our analysis above mainly focused on the washout from inverse decays, which matters when $N_{L,R}$ are still in the plasma. However, at $T\ll M_N$, we have to consider the $2\rightarrow2$ processes mediated by $N_{L,R}$. We first discuss reactions mediated by the sterile neutrinos in our single generation scenario. In addition, we also comment on the impact of introducing two heavier generations. Here, we only include the contribution from the off-shell sterile states, since the on-shell sterile neutrinos are already included in the decays and inverse decays.  

The relevant processes are the washout of the left-chiral lepton asymmetry from  $l_L l_L \leftrightarrow H_L^\dagger H_L^\dagger$ (and all other reactions related via crossing symmetry) together with the washout of the right-chiral asymmetry $l_R l_R \leftrightarrow H_R^\dagger H_R^\dagger$. As these processes do not depend on the presence of $N_{L,R}$ in the initial or final states, there is no Boltzmann suppression and one finds an exponential damping of the lepton asymmetries until the washout rates decouple \cite{Buchmuller:2004nz}. Additionally, we find that there is also the possibility of transmuting the left-chiral lepton asymmetry into a right-chiral asymmetry via, e.g. $l_L H_L \leftrightarrow l_R H_R$. For all three processes we find that the scattering rate from dimensional analysis scales as 
\begin{align}\label{eq:scat}
    \Gamma_\text{scat.} \simeq \left(\frac{8}{3}\right)^2 \left(\frac{m_{\nu_L}}{v_L^2}\right)^2 T^3\,,
\end{align}
because the exchange of the lightest $N_{L,R}$ sources $m_{\nu_L}$ via Eq.\eqref{eq:mnuL}.
To be conservative, we demand that 
\begin{align}    \frac{\Gamma_\text{scat.}}{H(T)}\Big|_{T=M_N} <1\,, 
\end{align}
because the leptonic asymmetries are produced between $M_N$ (strong washout from inverse decays) and $M_N/10$ (weak washout from inverse decays).  This condition implies the bound
\begin{align}\label{eq:22scat}
m_{\nu_L}< \SI{8.51e-2}{\electronvolt} \sqrt{\frac{\SI{e11}{\giga\electronvolt}}{M_N}}\,,
\end{align}
and is shown by the brown shaded region in Fig.~\ref{fig:scanpar1} and Fig.~\ref{fig:scanpar2}.

Estimating the impact of scattering processes mediated by the heavier generations, $N^{(2,3)}_{L,R}$, requires some care, because so far we have not specified how the heavier $N^{(2,3)}_{L,R}$ couple to the lightest $\nu_{L,R}$, that exclusively carry the leptonic asymmetry. We assume that $N_{L,R}^{(2,3)}$ couple dominantly to $\nu_{L,R}^{(2,3)}$ and only negligibly to $\nu_{L,R}$ with a small coupling $\tilde{Y}$ with  $\mathcal{O}(\tilde{Y}) \ll \mathcal{O}(Y)$.

 One can then expect a similar bound of 
 \begin{align}
     \tilde{Y} < 0.2 \sqrt{\frac{M_{N_{2,3}}}{10 M_N}} \left(\frac{\SI{e12}{\giga\electronvolt}}{M_N}\right)^\frac{1}{4}\,,
 \end{align}
which is automatically satisfied for our assumption $\tilde{Y}\ll Y$, and realistic values of $Y$. 

Furthermore, the neutral component of $H_R$ can condense, turning scattering processes into inverse decays.  The left-right conversion of neutrinos $H_L \leftrightarrow l_L \nu_R$ is known to be dangerous in scenarios that store equal and opposite chemical potentials in the left and right chiral neutrinos \cite{Dick:1999je}. To estimate the rate we can take the scattering rate in Eq.~\eqref{eq:scat}  for $l_L H_L \leftrightarrow l_R H_R$ and add the appropriate number of $v_R$ insertions
\begin{align}
    \Gamma_{L\leftrightarrow R} \simeq \left(\frac{8}{3}\right)^2 \left(\frac{m_{\nu_L}}{v_L^2}\right)^2 v_R^2 T\,.
\end{align}
The resulting ratio with the Hubble rate scales as $\Gamma_{L\leftrightarrow R}/H(T) \sim 1/T$ during radiation domination, thus, the process becomes more important with decreasing temperature. In order not to upset leptogenesis, we demand that it never thermalises before the decoupling of the $\text{SU}(2)_L$ sphaleron at around $T_c=\SI{130}{\giga\electronvolt}$ \cite{DOnofrio:2014rug}
\begin{align}
    \frac{\Gamma_{L\leftrightarrow R}}{H(T)}\Big|_{T=T_c} <1\,.
\end{align}
We obtain the bound
\begin{align}\label{eq:weak}
    m_{\nu_L}< \SI{3.1e-2}{\electronvolt} \sqrt{\frac{T_c}{\SI{130}{\giga\electronvolt}}} \left(\frac{\SI{e7}{\giga\electronvolt}}{v_R}\right)\,,
\end{align}
where we choose the largest possible $v_R$ allowed by the quality problem in Eq.~\eqref{eq:qual2} to consequently obtain the tightest limit. This bound is shown by the red shaded region in Fig.~\ref{fig:scanpar1} and Fig.~\ref{fig:scanpar2}. It can be seen that this bound is typically weaker than the one obtained from scattering in Eq.~\eqref{eq:22scat}, as our numerical analysis usually results in $M_N>\SI{e11}{\giga\electronvolt}$. Moreover, for lower values of $v_R$ allowed by the collider limit in Eq.~\eqref{eq:18}, the bound is clearly sub-leading as it scales with $1/v_R$. Therefore, this process does not obstruct low scale left-right symmetry breaking. 

\subsubsection{Gauge Boson Scatterings}
Leptogenesis in the minimal left-right symmetric model \cite{Pati:1974yy,Mohapatra:1974gc,Senjanovic:1975rk,Chang:1983fu,Chang:1984uy} usually proceeds via decays of the $\nu_R$ belonging to $l_R$ (see Ref.~\cite{Hati:2018tge} for a review), that obtains its mass from the VEV of an $\text{SU}(2)_R$ triplet via the Type II Seesaw mechanism \cite{Lazarides:1980nt,Schechter:1980gr,Mohapatra:1980yp,Cheng:1980qt,Wetterich:1981bx}. 
Gauge interactions of $\nu_R$ can   equilibrate it via processes like $e_R^\pm \nu_R \leftrightarrow e_R^\pm \nu_R$ via $W_R$ exchange, and the asymmetry stored in right chiral leptons can be washed out from processes like:
\begin{equation*}
 e_R^\pm W_R^\mp \leftrightarrow e_R^\mp W_R^\pm, \; \nu_R Z' \leftrightarrow \nu_R^\dagger Z',\; W_R^\pm W_R^\pm \leftrightarrow e_R^\pm e_R^\pm, \; Z' Z' \leftrightarrow \nu_R \nu_R \,,  
\end{equation*}
mediated by an insertion of the $\nu_R$ Majorana mass \cite{Ma:1998sq,Frere:2008ct} (see also Refs.~\cite{BhupalDev:2014hro,Gu:2017gra}). Equilibration via gauge interactions is irrelevant for the gauge singlets $N_{L,R}$, whose decays source the asymmetry in our set-up. 

The rate for the washout processes can be estimated via dimensional analysis for $T\gg m_{W_R} \gg m_{\nu_R}$, in terms of the $\text{SU}(2)_R$ gauge coupling $g_R$
\begin{align}
    \Gamma\left(e_R^\pm W_R^\mp \leftrightarrow e_R^\mp W_R^\pm,\; W_R^\pm W_R^\pm \leftrightarrow e_R^\pm e_R^\pm\right) \simeq g_R^4 \frac{m_{\nu_R}^2}{T}\,,
\end{align}
and it is evident that this process is suppressed compared to the usual left-right symmetric model due to the smallness of $m_{\nu_R}$ defined in Eq.~\eqref{eq:mnuR}. Similarly, for $\nu_R Z' \leftrightarrow \nu_R^\dagger Z', \; Z' Z' \leftrightarrow \nu_R \nu_R$, the same functional form applies, and the gauge coupling $g_R$ has to be replaced by the couplings to the $Z'$. 

Furthermore, these processes appear only after left-right symmetry breaking $T<v_R$ because $m_{\nu_R}\sim v_R^2/M_N$.
Thus, they become relevant only after the $\text{SU}(2)_R$ sphalerons have decoupled and $\Delta_{R}$ has been converted into a baryon asymmetry. Consequently, these processes cannot help with the cancellation between  $\Delta_{R}$ and $\Delta_{L}$ encountered in Section~\ref{sec:lowscale}.

Note that the above gauge scatterings could deplete $\Delta_{R}$ before the left-right equilibration between $\Delta_{L}$ and $\Delta_{R}$ (discussed above) can become important. However, since the bound from this process in Eq.~\eqref{eq:weak} was relatively weak to begin with, we do not consider this sequence of events further.

\section{Scenarios}\label{sec:scenarios}

As shown above, the total baryon asymmetry depends on whether leptogenesis occurs after or before the spontaneous breakdown of left-right symmetry at $T=v_R$. Below, we discuss the difference between the two scenarios. 

\subsection{Decays after High-Scale LRS Breaking $(M_N<v_R)$}

In this scenario, the resulting baryon asymmetry depends only on the asymmetry in the left-handed baryons produced by the $\text{SU}(2)_L$ sphaleron transitions \cite{Kuzmin:1985mm}
\begin{align}
    \Delta_B = \Delta_{B_L}  = c_\text{sph.}(T\ll v_R)\; \left(\kappa_L \varepsilon^L_L + \kappa_R \varepsilon^L_R \right) \;\frac{270 \zeta(3)}{8\pi^4 g_{*S}(T)}\,.
\end{align}
The sphaleron redistribution coefficient at $T< v_R$ only depends on the SM field content and their interactions, so the standard result applies, and the details can be found in Appendix \ref{app:belowLR}. The coefficient depends on which Yukawa interactions are equilibrated and therefore varies between the two limiting values \cite{Nardi:2005hs,Nardi:2006fx}
\begin{align}\label{eq:range}
    c_\text{sph.}(T\ll v_R) = 
    \begin{cases}
    \frac{28}{79}\quad &\text{for} \quad  \mathcal{O}(\SI{100}{\giga\electronvolt})\leq T\leq \SI{e5}{\giga\electronvolt}  \\
    \frac{2}{5}\quad &\text{for} \quad \SI{1.2e12}{\giga\electronvolt}\leq  T\leq \SI{2.5e12}{\giga\electronvolt}\,,
    \end{cases} 
\end{align}
and goes to zero above $\SI{2.5e12}{\giga\electronvolt}$ as the $\text{SU}(2)_L$ sphalerons drop out of equilibrium.

For our analytical estimates, we neglect $\delta_{L,R}$ and introduce the effective amount of $CP$ violation
\begin{align}
\varepsilon_\text{eff.} \equiv  \varepsilon_L^L +  d(r,K_L) \varepsilon_R^L\,,
\end{align}
in terms of the ratio of efficiencies
\begin{align}
    d(r,K_L) \equiv \frac{\kappa_R}{\kappa_L} \simeq 
    \begin{cases}
    {1}/{r}\quad &\text{for}\quad K_L\ll1\,,\\
    {K_L^3}/{r} \quad &\text{for} \quad 1<K_L<\sqrt{r}\\
    \sqrt{r} \quad &\text{for}\quad  \sqrt{r}<K_L
    \end{cases}\,.
\end{align}
In the first (last) line both $N_{L,R}$ are in the weak (strong) washout regime and in the second line $N_L$ is in the strong washout regime, whereas $N_R$ is in the weak regime.
This allows us to determine the following approximate expressions for
\begin{align}
   |\varepsilon_\text{eff.}|\simeq \frac{Y^2 \sin{(2\varphi)}}{4\pi} j(r,K_L)\,,
\end{align}
where for  $r\gtrsim 1$
\begin{align}\label{eq:line1}
j(r,K_L)=\frac{1}{r-1}
\begin{cases}
    1 \quad &\text{for}\quad  K_L\ll1,\; \sqrt{r}<K_L\\
   K_L^3\quad &\text{for} \quad  \quad 1<K_L<\sqrt{r}
\end{cases}\,,
\end{align}
and for $r\gg1$
\begin{align}\label{eq:line2}
j(r,K_L) = 
\begin{cases}
    {5}/{(8 \sqrt{r})}&\text{for}\quad  K_L\ll1,\; 1<K_L<\sqrt{r}\\
    {(\log{(r)} -3)}/{4} \quad &\text{for} \quad  \sqrt{r}<K_L
\end{cases}\,.
\end{align}
\begin{figure}[!t]
    \centering
    \includegraphics[width=0.49\linewidth]{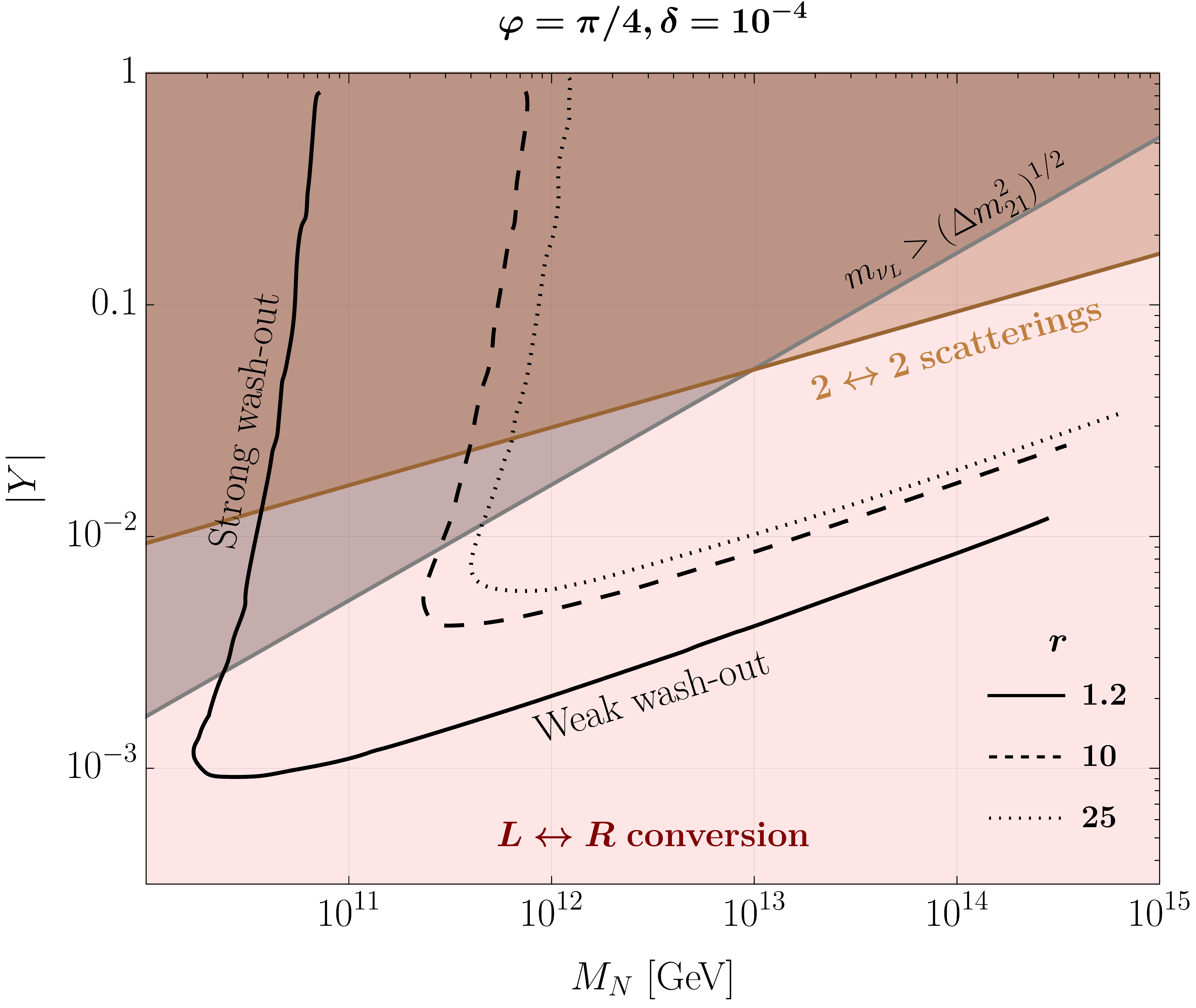}
    \includegraphics[width=0.49\linewidth]{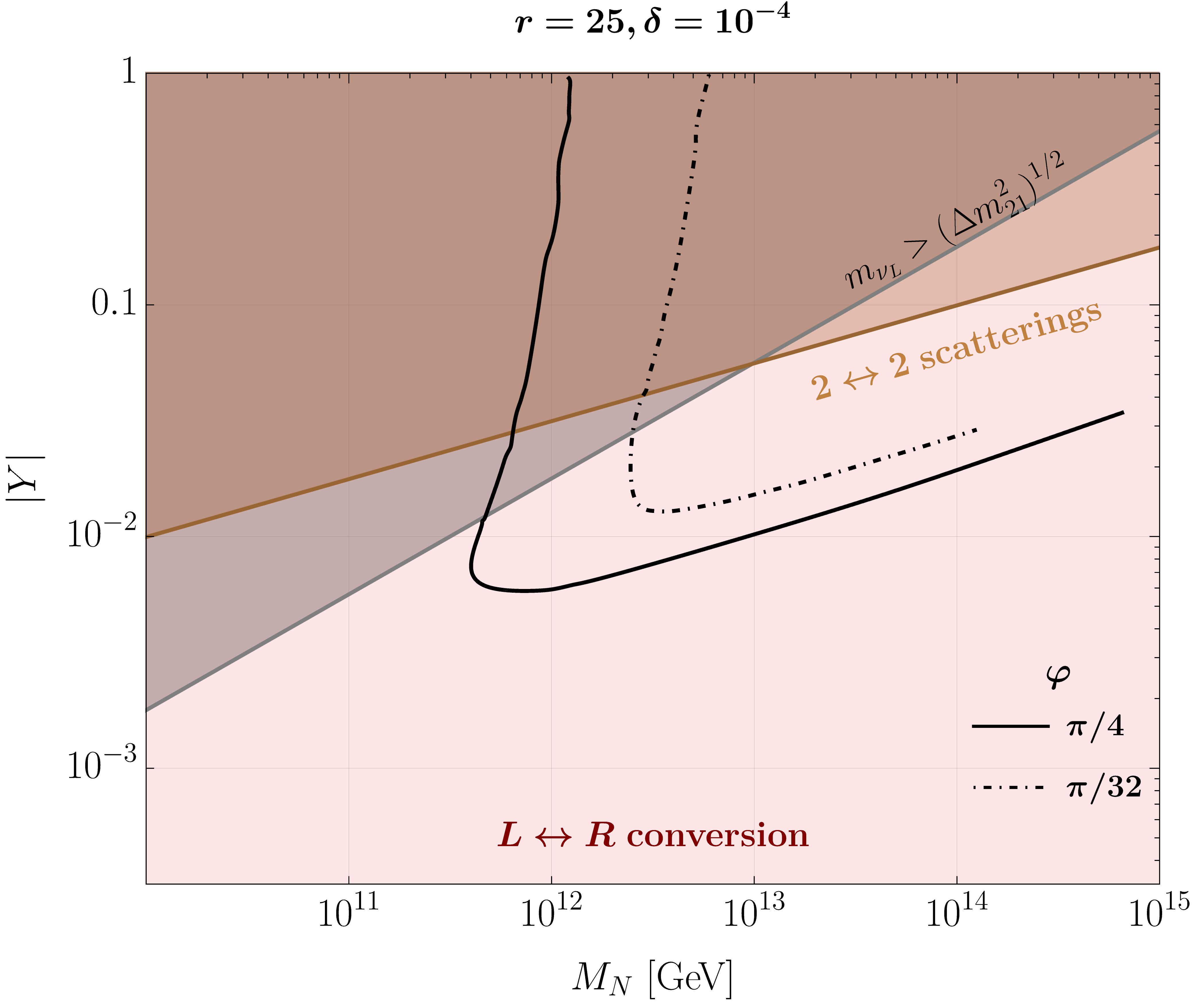}
    \caption{Contours of the observed baryon asymmetry (95\% CL) in the $M_N$ - $|Y|$ plane for decays after the high LRS breaking. \textit{Left:} Varying $r$ with fixed $\delta$ and $\varphi = \pi/4$. \textit{Right:} Varying $\varphi$ for a fixed value of $r$ and $\delta$. In the shaded gray region, the lightest neutrino mass is larger than the solar mass splitting $\sqrt{\Delta m_{21}^2}$, whereas the shaded light brown region is excluded from $2 \leftrightarrow 2$ scatterings (see Eq.~\eqref{eq:22scat}). Including $L \leftrightarrow R$ conversion for {$v_R =10 M_N$} excludes the entire viable region (light red shading, see Eq.~\eqref{eq:weak}). The left and the right wings of all the boomerang shaped contours correspond to the strong and weak washout regimes, respectively.}
    \label{fig:scanpar1}
\end{figure}
One might worry that $|\varepsilon_\text{eff.}|$ grows logarithmically with $r$ for $\sqrt{r}<K_L$, but in practice we find that for very large $r$ that $K_L$ is typically not large enough to establish the strong washout regime for $N_R$. For the analytical estimate, we continue with $r=10$ and the expression in the first line of Eq.~\eqref{eq:line2}. The estimate for the baryon asymmetry reads for $\varphi=\pi/3$
\begin{align}\label{eq:asym1}
    |\Delta_B| \simeq 8.76\times 10^{-11} \left(\frac{\sin{\left(2\varphi\right)}}{0.87}\right) \sqrt{\frac{10}{r}}   \begin{cases}
        \left(\frac{\SI{2.5e14}{\giga\electronvolt}}{M_N }\right)^2 \left(\frac{Y}{0.02}\right)^6 \quad &\text{for} \quad K_L=0.1\\
        \left(\frac{M_N}{\SI{2.6e11}{\giga\electronvolt}}\right) \quad &\text{for} \quad K_L=3
    \end{cases}\,.
\end{align}
One can see that the result for the strong washout regime does not explicitly depend on $Y$, and we find for the chosen values of $M_N$ and $K_L$: $Y=3.7\times10^{-3}$.

The points in the $M_N - |Y|$ plane that reproduce the observed baryon asymmetry, obtained from the numerical solution of the Boltzmann equations of Eq.~\eqref{eq:beq} are shown in Fig.~\ref{fig:scanpar1} for different values of $r$ and $\varphi$, along with the constraints from washout processes discussed above. We find that $\delta$ has the least impact on the preferred parameter space, and fix $\delta = 10^{-4}$. The constraint on the lightest neutrino being effectively massless cuts away most of the parameter space for the strong washout regime. Fig.~\ref{fig:scanpar1} also evinces that leptogenesis is possible for 
\begin{align}
    M_N > \mathcal{O}(\SI{e10}{\giga\electronvolt})\,.
\end{align}
As we assume that $v_R >M_N$ in this scenario, the scalar and the gauge bosons of the left right symmetric model are not accessible at any terrestrial experiments. Furthermore, such large $v_R$ might violate the bound from the quality problem in Eq.~\eqref{eq:qual2}, when treating generalized parity as a spontaneously broken gauge symmetry. Saving the parity-based solution to the strong $CP$ problem then requires a mechanism to suppress the coefficients of the putative higher dimensional operators.

{For the depicted range of $v_R>M_N>\SI{e10}{
\giga\electronvolt}$ the limit from Eq.~\eqref{eq:weak}, which reads $m_{\nu_L}< \SI{3.1e-5}{\electronvolt}$, rules out all of the available parameter space as can be seen in Fig.~\ref{fig:scanpar1}, so the regime $v_R>M_N$ is not viable for Leptogenesis. }

\subsection{Decays before Low-Scale LRS Breaking $(M_N>v_R)$}\label{sec:lowscale}

Since the decays of $N_{L,R}$ each produce asymmetries in both $l_{L,R}$, and we expect both  $\text{SU}(2)_L$ and  $\text{SU}(2)_R$ sphalerons  \cite{Kuzmin:1985mm} to be in equilibrium in this scenario, we have to add the resulting asymmetries of the left- and right-chiral baryons, weighted by their opposite $B-L$ charges (see Table~\ref{tab:fields})  
\begin{align}\label{eq:yban}
    \Delta_B = \Delta_{B_L} - \Delta_{B_R} = c_\text{sph.}(T\gg v_R)\; \left( \kappa_L \left(\varepsilon^L_L - \varepsilon^R_L\right) + \kappa_R \left( \varepsilon^L_R - \varepsilon^R_R \right) \right) \frac{270 \zeta(3)}{8\pi^4 g_{S}(T)}\,.
\end{align}
The sphaleron redistribution coefficient reads 
\begin{align}\label{eq:csph}
     c_\text{sph.}(T\gg v_R)=\frac{28}{79}\,,
\end{align}
where we assume generation independent chemical potentials for each species and equilibration of all Yukawa interactions. The details for this computation can be found in Appendix \ref{app:aboveLR}. Though this coefficient has a mild temperature dependence similar to Eq.~\eqref{eq:range}, we find its impact to be numerically sub-leading. In order to obtain an analytical estimate, we define 
\begin{align}
\varepsilon_\text{eff.} \equiv \varepsilon_L^L -\varepsilon_L^R + d(r,K_L)\left(\varepsilon_R^L-\varepsilon_R^R\right)\,.
\end{align}
We find the following limiting cases for the effective CP asymmetry produced per decay
\begin{align}\label{eq:sym}
|\varepsilon_\text{eff.}| \simeq \frac{Y^2}{4\pi}\sin{(2\varphi)}\;  j(r,K_L,\delta_L)\,,
\end{align}
where for  $r\gtrsim 1$
\begin{align} 
j(r,K_L,\delta_L)=\frac{\delta_L}{r-1}
\begin{cases}
    1 \quad &\text{for}\quad  K_L\ll1,\; \sqrt{r}<K_L\\
  {(1+ K_L^3)}/{2}\quad &\text{for} \quad  1<K_L<\sqrt{r}
\end{cases}\,,
\end{align}
and for $r\gg1$
\begin{align} 
j(r,K_L,\delta_L) = \frac{3\delta_L}{4\sqrt{r}}\,.
\end{align}
The behavior of $\varepsilon_\text{eff.}$ can be understood as follows: In the regime of exact parity, we have equal masses for $N_{L,R}$ ($r=1$) and the $CP$ asymmetry in the decays of $N_L$ and $N_R$ to the same final states cancel out, because for $d(r,K_L)=1$, we have $\varepsilon_\text{eff.} = \varepsilon_L^L +\varepsilon_R^L -\varepsilon_L^R-\varepsilon_R^R=0$ (see Eq.~\eqref{eq:Weinberg}). This further illustrates why a soft-breaking in the sterile neutrino sector ($r\neq 1$) is necessary. 

To understand why the factor of $\delta_L$ appears, note that we require $\varepsilon_L^L-\varepsilon_L^R\neq 0 \neq \varepsilon_R^L-\varepsilon_R^R$.
By comparing Eq.~\eqref{eq:cpasym1} with Eq.~\eqref{eq:cpasym2}, one can see that the previous conditions amount to $f(r)\neq f^{(2)}(r,\delta_L)$ and $f(1/r)\neq f^{(2)}(1/r,\delta_L/r)$, which necessitates $\delta_L\neq 0$. Since $\delta_L\propto m_{H_R}^2$, this is nothing more than the soft breaking of parity in the Higgs sector. At $T>v_R$, the VEV and therefore the mass of $H_R$ encoded in $\delta_L$ vanishes and consequently we have $\varepsilon_\text{eff.}=0$. Therefore, this calls for a remedy if we still want leptogenesis to occur above $v_R$. In the following, we break the parity symmetry explicitly at the level of the Yukawas by replacing $Y\rightarrow Z$ in Eq.~\eqref{eq:41}.   We take the generalized parity  to be only approximate $Z_{RR}\simeq  Z_{LL}^*$ and proceed with a single generation estimate
\begin{align}
    Z_{LL} &= Y_{LL} + i X, \quad 
    Z_{RR} = Y_{LL}^* + i X^*,\\
    Z_{LR} &= Y_{LR},\quad \quad \quad
    Z_{RL} =Y_{LR}^*\,.
\end{align}
where we define 
\begin{align}
  X=  Y\; \omega\; e^{i \frac{\chi}{2}}, \quad |\omega|\ll1\,,
\end{align}
in terms of $Y, \omega, \chi  \in \mathbb{R}$.
One can understand $X$ as a small perturbation to the previously introduced Yukawas due to the choice of $|\omega|\ll1$ and consequently the order of magnitude of the neutrino masses from Section~\ref{sec:numass} will not change. 
We still assume that $Y_{RL}=Y_{LR}^*$, because if we were to shift all couplings with the same $\omega, \chi$; we would obtain parity violating couplings that are pairwise identical $Z_{LR}= Z_{RR}$ and $Z_{RL}=Z_{LL}$, leading to 
$\varepsilon_L^L-\varepsilon_L^R= 0 = \varepsilon_R^L-\varepsilon_R^R$. 
It would be interesting to explain the explicit parity breaking by radiative corrections, which we leave for future investigation.
Alternatively as discussed in Section~\ref{sec:qual}, such couplings can arise from higher dimensional operators after parity is spontaneously broken by the VEV of the singlet scalar $S$
\begin{align}\label{eq:SSB2}
        \frac{i S}{\Lambda_{\rm UV}} \left( \gamma_{LR} l_L \varepsilon H_L N_R + \gamma_{LR}^* l_R \varepsilon H_R N_L +  
        \gamma_{LL} l_L \varepsilon H_L N_L   + \gamma_{LL}^* l_R \varepsilon H_R N_R\right) + \text{H.c.}\,,
\end{align}
where we assume that only one coupling is important 
\begin{align}
    |\gamma_{LR}| \ll |\gamma_{LL}|\,,
\end{align}
and identify
\begin{align}
    X= \frac{v_S}{\Lambda_{\rm{UV}}} \gamma_{LL}\,.
\end{align}
This particular realization requires the following  hierarchy of scales 
\begin{align}
  \Lambda_{\rm{UV}} > v_S \geq M_{RR}>M_N >v_R\,,
\end{align}
which could have implications for a putative UV completion.
For the above choice of parity breaking, one can deduce that
\begin{align}
   \varepsilon^L_L \equiv    \varepsilon_L(N_L) &=
   \frac{ \text{Im}\left(\left(Z_{LL}Z_{LR}^*\right)^2\right) f\left(r\right)  +
   \text{Im}\left(Z_{LR}^* Z_{RL} Z_{LL} Z_{RR}^*\right) f^{(1)}\left(r,0\right)}{8\pi (|Z_{RL}|^2 +|Z_{LL}|^2)}\,,\nonumber\\
    \varepsilon^L_R  \equiv  \varepsilon_L(N_R) &=
    -\frac{ \text{Im}\left(\left(Z_{LL}Z_{LR}^*\right)^2\right) f\left(\frac{1}{r}\right)  +
   \text{Im}\left(Z_{LR}^* Z_{RL} Z_{LL} Z_{RR}^*\right) f^{(1)}\left(\frac{1}{r},0\right)}{8\pi (|Z_{LR}|^2+|Z_{RR}|^2}\,, 
\end{align}
and furthermore 
\begin{align} 
    \varepsilon^R_L \equiv \varepsilon_R(N_L) &=
   \frac{ \text{Im}\left(\left(Z_{RL} Z_{RR}^*\right)^2\right) f\left(r\right)  +
   \text{Im}\left(Z_{LR}^* Z_{RL} Z_{LL} Z_{RR}^*\right) f^{(1)}\left(r,0\right)}{8\pi (|Z_{RL}|^2 +|Z_{LL}|^2)}\,,\nonumber\\
    \varepsilon^R_R \equiv  \varepsilon_R(N_R) &=
    -\frac{\text{Im}\left(\left(Z_{RL} Z_{RR}^*\right)^2\right) f\left(\frac{1}{r}\right)  +
   \text{Im}\left(Z_{LR}^* Z_{RL} Z_{LL} Z_{RR}^*\right) f^{(1)}\left(\frac{1}{r},0\right)}{8\pi (|Z_{LR}|^2+|Z_{RR}|^2}\,.
\end{align}

Since the couplings of $N_L$ and $N_R$ are not identical anymore, we find that the relation in Eq.~\eqref{eq:Weinberg} no longer holds. Due to the modified couplings we find that e.g. $\text{Im}\left(\left(Z_{LL}Z_{LR}^*\right)^2\right)\neq \text{Im}\left(\left(Z_{RL} Z_{RR}^*\right)^2\right)$, thus, $\varepsilon_L^L - \varepsilon_L^R\neq 0$. These new couplings  modify the amount of $CP$ asymmetry as follows 
\begin{align}
|\varepsilon_\text{eff.}| \simeq \frac{Y^2}{4\pi}\cos{\left(\frac{3\varphi+\chi}{2}\right)}\;  j(r,K_L,\omega)\,,
\end{align}
where for  $r\gtrsim 1$
\begin{align}
j(r,K_L,\omega)=\frac{2\omega}{(r-1)}
\begin{cases}
    1 \quad &\text{for}\quad  K_L\ll1,\; \sqrt{r}<K_L\\
   {(1+K_L^3)}/{2}\quad &\text{for} \quad  \quad 1<K_L<\sqrt{r}
\end{cases}\,,
\end{align}
and for $r\gg1$
\begin{align}
j(r,K_L,\omega)=\frac{3\omega}{2} 
\begin{cases}
    {1}/{\sqrt{r}}&\text{for}\quad  K_L\ll1,\; 1<K_L<\sqrt{r}\\
    {(2(\log{(r)} -2))}/{3} \quad &\text{for} \quad  \sqrt{r}<K_L
\end{cases}\,.
\end{align}
The dependence on $\omega \ll 1$ is now explicit and consequently our scenario features a novel source of asymmetry suppression not present in the usual Type I Seesaw leptogenesis. To minimize the number of free parameters, we will take $\chi=0$ in the analysis. Note that the functional dependence on $\varphi$ has changed compared to the case with parity-symmetric Yukawas in Eq.~\eqref{eq:sym}. 

With these ingredients we can estimate the resulting baryon asymmetry for $\varphi=\pi/4$ to be 
\begin{align}
    |\Delta_B| \simeq 8.76\times 10^{-11} \left(\frac{\cos{\left(\frac{3\varphi}{2}\right)} }{0.38}\right) \left(\frac{\omega}{0.1}\right)\sqrt{\frac{10}{r}}  \begin{cases}
        \left(\frac{\SI{2.4e15}{\giga\electronvolt}}{M_N}\right)^2 \left(\frac{Y}{0.06}\right)^6 \quad &\text{for} \quad K_L=0.1\\
        \left(\frac{M_N}{\SI{2.4e12}{\giga\electronvolt}}\right) \quad &\text{for} \quad K_L=3
    \end{cases}\,.
\end{align}
The baryon asymmetry in the strong washout regime is still independent of $Y$  and we take $Y=0.011$ to obtain $K_L=3$ for the above value of $M_N$.
\begin{figure}[!t]
    \centering
    \includegraphics[width=0.49\linewidth]{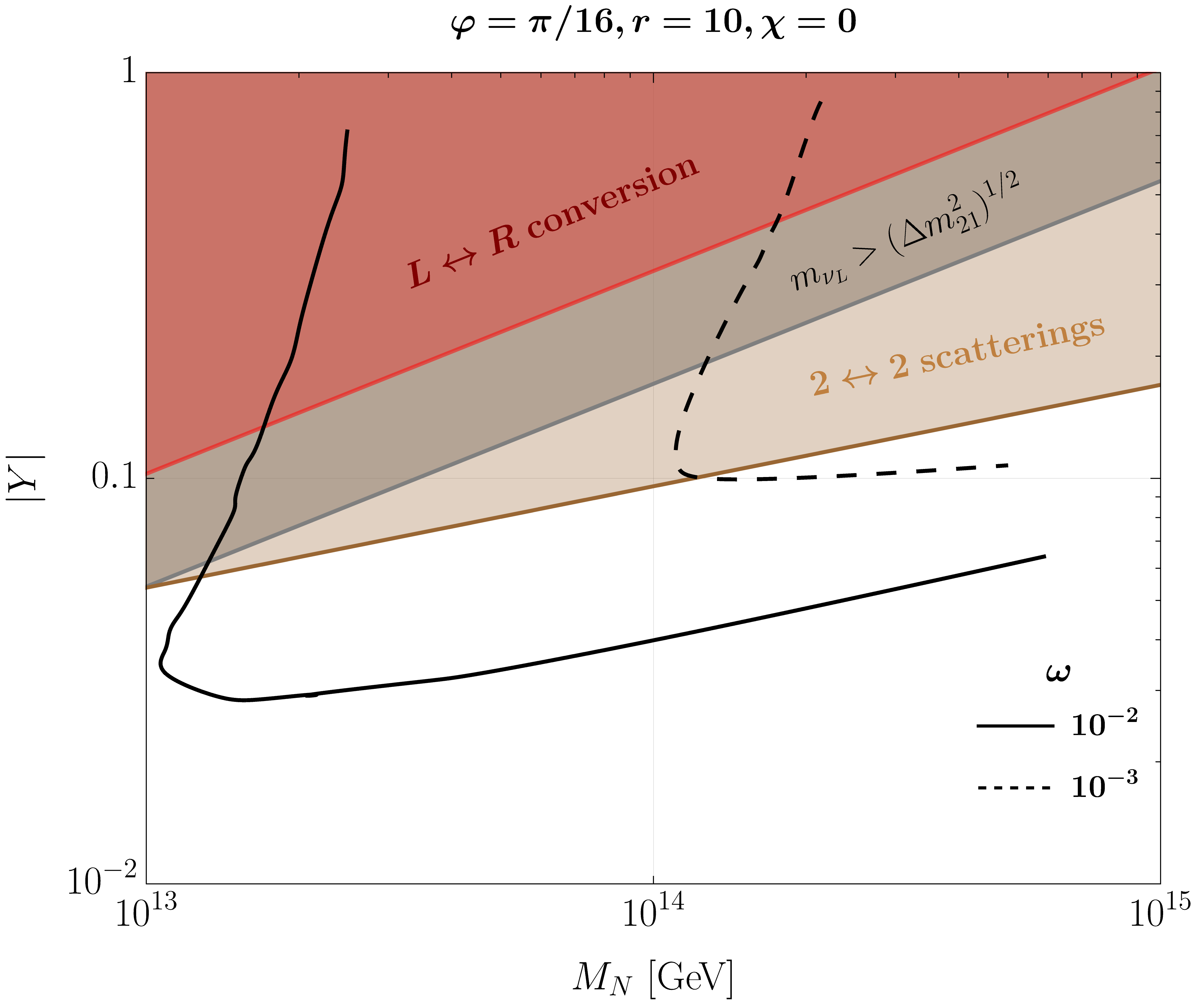}
    \includegraphics[width=0.49\linewidth]{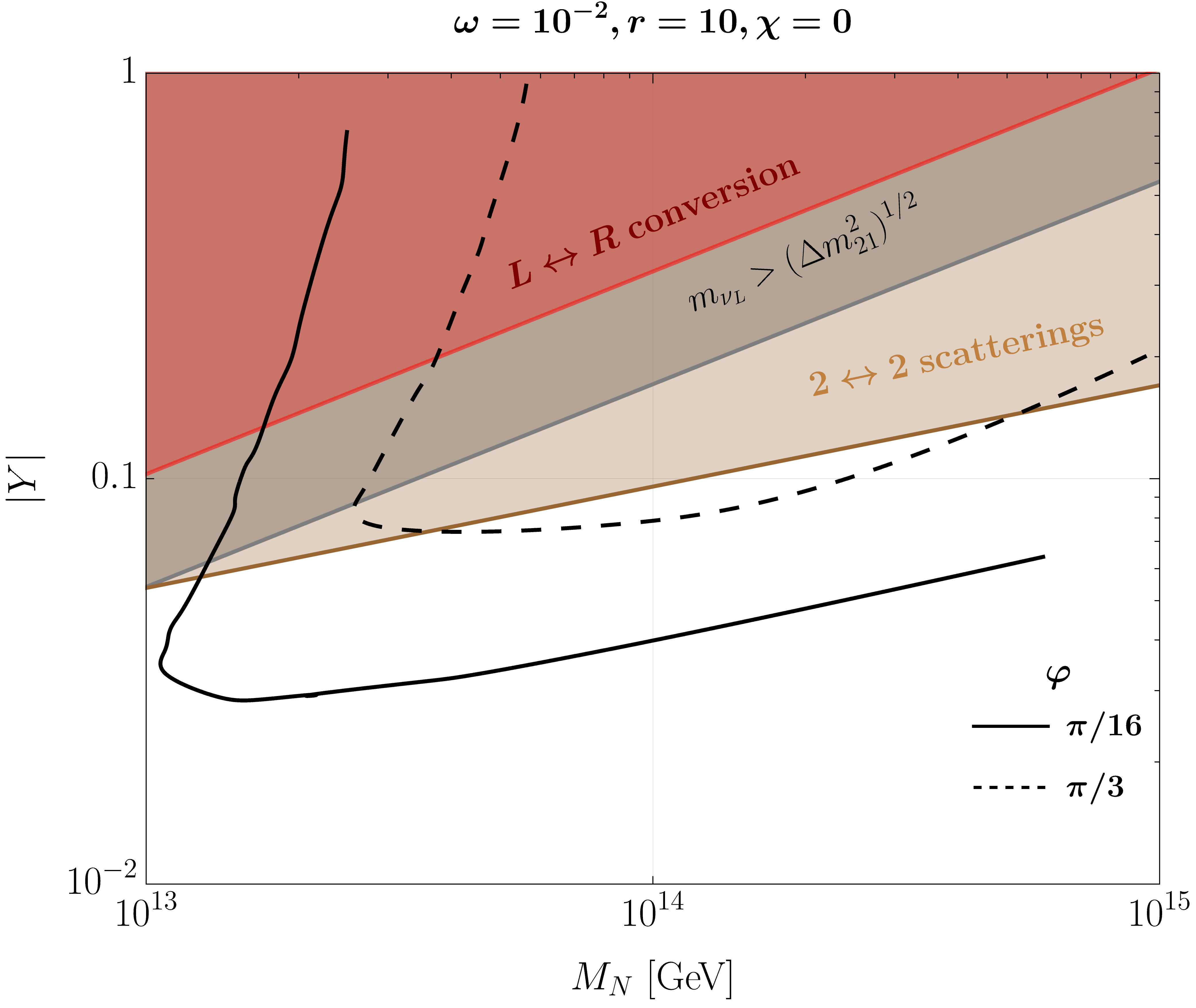}
    \caption{Same as Fig.~\ref{fig:scanpar1}, but for decays before LRS breaking. \textit{Left:} Varying $\omega$ with fixed $r$ and $\varphi$, and setting $\chi=0$. \textit{Right:} Varying $\varphi$ with fixed $r$ and $\omega$. The shaded regions correspond to the strongest constraints, obtained for $\varphi=\pi/16$.}
    \label{fig:scanpar2}
\end{figure}

As anticipated, the presence of $\omega\ll1$ pushes the required values of $M_N$ to higher masses when compared with Eq.~\eqref{eq:asym1}. The result of a numerical scan involving the solutions to the Boltzmann equations in section \ref{sec:BE} is shown in Fig.~\ref{fig:scanpar2} for various choices of $r$, $\varphi$ and $\omega$. For $\omega<0.1$ we find that 
\begin{align}
    M_N > \mathcal{O}(\SI{e12}{\giga\electronvolt})\,,
\end{align}
in order to reproduce the observed baryon asymmetry, and this agrees with our analytical expectation. The price for leptogenesis with $M_N>v_R$ is both a larger $M_N$ and the need for a breaking of parity at the level of the Yukawa couplings to $N_{L,R}$. This scenario is compatible with the bound on $v_R$ from the quality problem of the solution to the strong $CP$ problem in Eq.~\eqref{eq:qual2}. Moreover, as this scenario does not constrain $v_R$ directly, low-scale left-right symmetry breaking in reach of next generation colliders can be realized. 

Note that since we consider $T_\text{RH}>M_{RR}>M_N>v_R$ the spontaneous breaking of gauged $\mathcal{P}$ is accompanied by dangerous domain walls as explained in Section~\ref{sec:DW}, which are not inflated away. Hence, this scenario prefers $\mathcal{P}$ realized as a global symmetry.

Another variant of our idea is to leave the parity relation between the couplings intact and introduce an additional parity-breaking coupling specific to $l_R$ instead. This can lead to a selective wash-out of the right chiral lepton asymmetry, which is sketched in Appendix \ref{sec:WObreak}.

\section{Light Right-Handed Neutrino Cosmology}\label{sec:RHN}

In this section, we discuss the productions mechanisms and cosmological implications of the light RHN in the model.\\ 

\noindent\textbf{Production:} The neutral components of the right-handed lepton doublet $l_R$ are thermally produced via their $\text{SU}(2)_R$ gauge interactions and their scattering rate is given by 
\begin{align}
    \Gamma_{\nu_R} = \left(\frac{v_R}{v_L}\right)^4 \Gamma_{\nu_L}\,,
\end{align}
where $\Gamma_{\nu_L}\simeq T^5/v_L^4$ is the left-handed neutrino scattering rate from the Fermi interaction at $T\ll m_W$, which freezes out at $T_{\nu_L}^\text{FO}\simeq\SI{1}{\mega\electronvolt}$, right at the onset of BBN. The corresponding decoupling temperature for $\nu_R$ reads
\begin{align}\label{eq:T}
    T_{\nu_R}^\text{FO}\simeq\SI{1}{\mega\electronvolt} \left(\frac{v_R}{v_L}\right)^\frac{4}{3}\,,
\end{align}
and we see that it is entirely determined by $v_R$.

One may wonder about the relevance of active to sterile oscillations, which for $m_{\nu_{R}}\gg m_{\nu_{L}}$ can be analytically  computed following Refs.~\cite{Notzold:1987ik,Boyarsky:2009ix}. Here one considers the production rate of a single $\nu_R$ via its mixing with $\nu_L$ given by Eq.~\eqref{eq:mix}, which up to thermal corrections reads
\begin{align}
    \Gamma_\text{mix} \simeq \frac{1}{4}\left(\frac{v_L}{v_R}\right)^4 \Gamma_{\nu_L}\,.
\end{align}
It is evident that this channel is subdominant to the production from right-chiral gauge currents
\begin{align}
    \frac{\Gamma_\text{mix} }{\Gamma_{\nu_R}}\simeq \frac{1}{4} \left(\frac{v_L}{v_R}\right)^8 \ll1\,.
\end{align}
\noindent\textbf{Relic abundance:} We find that the lightest $\nu_R$ is always relativistic at decoupling and the energy density for such a relic reads
\begin{align}
    \frac{\rho_{\nu_R}}{s}= m_{\nu_R} \frac{135 \zeta(3)}{4\pi^4 g_S(T_{\nu_R}^\text{FO})} = \SI{0.44}{\electronvolt} \left(\frac{m_{\nu_R}}{\SI{111}{\electronvolt}}\right)\left(\frac{106.75}{g_S(T_{\nu_R}^\text{FO})}\right)\,,
\end{align}
where $\rho_\text{DM}/s=\SI{0.44}{\electronvolt}$ corresponds to the observed dark matter relic abundance of $h^2 \Omega_\text{DM} = h^2 (\rho_\text{DM}/\rho_\text{crit}) (s_0/s)=0.12$ \cite{Planck:2018vyg} in terms of the critical density $\rho_\text{crit}$ and the present day entropy density $s_0$. 

The value of $m_{\nu_R}$ could increase by less than a factor of two, if the decoupling from the plasma happened so early that some additional particles present in the Universal Seesaw were still relativistic, so that $g_S(T_{\nu_R})$ is larger than its maximum value in the Standard Model of 106.75.\footnote{The largest possible number of relativistic degrees of freedom is obtained when all particles except $N_{L,R}$ behave like radiation, and it reads
$g_{S,\text{max}}=192.5$.}
Significantly larger $m_{\nu_R}$ can be accommodated if the relic density is suppressed by the entropy release from a long-lived particle, leading to an epoch of early matter domination. We will comment on this possibility at the end of this section.

We find that the RHN mass is in general too small to comply with the Tremaine-Gunn bound \cite{Tremaine:1979we} derived from the composition of the dark matter halos using the Pauli exclusion principle. A recent reevaluation using data from dwarf spheroidal galaxies  finds that the thermal relic dark matter mass should be \cite{Bezrukov:2025ttd}
\begin{align}
    m_{\nu_R} > (280-490)\;\text{eV}.
\end{align}
We estimate the $\nu_R$ velocity following Ref.~\cite{Acero:2008rh}
\begin{align}
    v_{\nu_R}(z) = 3.151 (1+z) \frac{T_{\nu_L,0}}{m_{\nu_R}} \left(\frac{10.75}{g_S(T_{\nu_R}^\text{FO}}\right)^\frac{1}{3}\,,
\end{align}
in terms of the redshift $z$ and the present day neutrino temperature $T_{\nu_L,0}= (4/11)^\frac{1}{3} T_\text{CMB}$, where $T_\text{CMB}= 2.726\;\text{K}$ is the present day CMB temperature. 
At the time of CMB decoupling ($z\simeq 1100$) we obtain
\begin{align}
      v_{\nu_R,\text{CMB}}\simeq 0.27 \left(\frac{\SI{1}{\electronvolt}}{m_{\nu_R}}\right) \left(\frac{106.75}{g_S(T_{\nu_R}^\text{FO})}\right)^\frac{1}{3}\,,
\end{align}
from  which it is evident that the $\nu_R$ population will typically propagate with ultra-relativistic speeds during the onset of structure formation. Cold dark matter has a velocity that is negligible at the time of structure formation. Relics with a velocity one or two orders of magnitude below the speed of light are usually referred to as hot dark matter \cite{Viel:2005qj} and relics with  velocities between these two extremes are known as warm dark matter. Therefore, the choice of $m_{\nu_R}$ mostly determines whether $\nu_R$ falls into the hot ($m_{\nu_R}\lesssim \mathcal{O}(\SI{100}{\electronvolt})$) or warm  ($m_{\nu_R}\gtrsim \mathcal{O}(\SI{100}{\electronvolt})$) dark matter regime. 

Since the relativistic $\nu_R$ are free-streaming after their freeze-out, they can escape gravitational potential wells and lead to the washout of structures below their mean free path. This delay in structure formation is heavily constrained by observations, which set a limit on the dark matter velocity, from which a limit on $m_{\nu_R}$ can be derived. Data from the ionization history of our universe \cite{Lopez-Honorez:2017csg} 
and observations of the Lyman-$\alpha$ forest \cite{Irsic:2017ixq} 
constrain the velocity to be $v_{\nu_R}<\mathcal{O}(10^{-4})$. The analysis of Lyman-$\alpha$ data of Ref.~\cite{Viel:2013fqw}
allows 
\begin{align}
    m_{\nu_R} >\SI{3.3}{\kilo\electronvolt}.
\end{align}
Data from 21-cm lines will be able to probe  $v_{\nu_R}\gtrsim \mathcal{O}(10^{-5})$ \cite{Sitwell:2013fpa} and could  strengthen  this limit  to \cite{Munoz:2019hjh}
\begin{align}
    m_{\nu_R} >  \SI{14}{\kilo\electronvolt}\,.
\end{align}
As we show that we are typically in the sub-keV regime, this implies that the lightest $\nu_R$ without entropy dilution can not account for all the observed dark matter. However, this still leaves the option that it only contributes a fraction of the relic density
\begin{align}
    f_{\nu_R} = \frac{\Omega_{\nu_R}}{\Omega_{\nu_R}+ \Omega_\text{DM}}\,.
\end{align}
Lyman-$\alpha$ data constrains the fraction of warm dark matter  to be $\mathcal{O}(30\%)$ \cite{Boyarsky:2008xj}. A recent analysis of hot dark matter using \verb|Planck| CMB temperature and polarization data together with weak gravitational lensing data from \verb|KiDS| in Ref.~\cite{Peters:2023asu}  found that $f_{\nu_R}<8\%$ for $m_{\nu_R}\leq \SI{20}{\electronvolt}$ and $f_{\nu_R}<25\%$ for $m_{\nu_R}\leq \SI{80}{\electronvolt}$. This particular analysis can not place bounds for $m_{\nu_R}>\SI{200}{\electronvolt}$. To be conservative, we will demand that
\begin{align}\label{eq:frac}
    f_{\nu_R}<10\%\,,
\end{align}
independently of the precise value of $m_{\nu_R}$.
Ref.~\cite{Xu:2021rwg} analyzed  CMB, gravitational lensing and \texttt{BOSS} galaxy power-spectrum  data to set a limit of 
\begin{align}
    m_{\nu_R}<\SI{2.3}{\electronvolt}\,,
\end{align}
for relics that decouple above the electroweak transition, which implies $v_R>\SI{2.7e6}{\giga\electronvolt}$ for our case. For later decoupling (smaller $v_R$), the limit on the mass becomes stronger and no limit can be set for $m_{\nu_R}>\SI{10}{\electronvolt}$.

\noindent\textbf{Constraints from $\Delta N_{\text{eff.}}$:} The single generation of sub-keV $\nu_R$ will be relativistic at least during BBN, and therefore it will contribute to the dark radiation abundance encoded in the shift of the effective number of neutrinos $N_\text{eff.}$ 
\begin{align}
    \Delta N_\text{eff.} \simeq 0.05 \left(\frac{106.75}{g_S(T_{\nu_R}^\text{FO})}\right)^\frac{4}{3}\,.
\end{align}
The observed abundances of the light elements produced during BBN constrain $\Delta N_\text{eff.} <0.3$ \cite{Fields:2019pfx,Yeh:2022heq}.
This implies a decoupling temperature larger than about 160 MeV, which via Eq.~\eqref{eq:T} corresponds to $v_R > \SI{11}{\tera\electronvolt}$. If $\nu_R$ is relativistic at CMB decoupling, the combined bound from BBN and CMB is $\Delta N_\text{eff.} <0.14$ \cite{Fields:2019pfx,Yeh:2022heq} and this enforces 
\begin{align}
    v_R > \SI{12.4}{\tera\electronvolt}\,,
\end{align}
for a single generation of $\nu_R$. This limit is  weaker than the strongest collider limit of Eq.~\eqref{eq:18}. If there are two light generations of $\nu_R$ present in the plasma, the limit increases to $\SI{90}{\tera\electronvolt}$. We do not include the bound from the latest \verb|SPT| data \cite{SPT-3G:2025bzu}, because it demands that $\Delta N_\text{eff.}<0$.

Next generation CMB experiments such as  \verb|CMB-S4|  \cite{Abazajian:2019eic,annurev-nucl-102014-021908} or  \verb|PICO|  \cite{NASAPICO:2019thw} are expected to probe down to $\Delta N_\text{eff.} <0.06$, which would  amount to a limit of 
\begin{align}
    v_R >\SI{400}{\tera\electronvolt}\,.
\end{align}

\noindent\textbf{CMB limits:} Whether $\nu_R$ is relativistic at the time of CMB decoupling ($T\simeq \SI{1}{\electronvolt}$) depends on its mass; for non-relativistic $\nu_R$ it is customary to constrain the effective neutrino mass \cite{Feng:2017nss,Planck:2018vyg}
 \begin{align}
     m_{\nu_R}^\text{eff.} = \SI{94.1}{\electronvolt} h^2 \Omega_{\nu_R}= \SI{11.3}{\electronvolt} \left(\frac{\rho_{\nu_R}/s}{\SI{0.44}{\electronvolt}}\right)\,.
 \end{align}
The precise limit on this quantity depends on the combination of datasets and priors. Here, we adopt the limit from Planck data together with gravitational lensing and baryon acoustic oscillations of \cite{Planck:2018vyg}
\begin{align}\label{eq:meff}
    m_{\nu_R}^\text{eff.} < \SI{0.65}{\electronvolt}\,,
\end{align}
which gives a bound of similar strength compared to our limit on the fraction of the relic abundance in Eq.~\eqref{eq:frac}.
 
The CMB bounds that were computed  in Refs.~\cite{Vincent:2014rja,Bridle:2016isd} and compiled in Ref.~\cite{Bolton:2019pcu} turn out to be sub-leading compared to limits on the relic density in Eq.~\eqref{eq:frac} and the collider constraint in Eq.~\eqref{eq:18}. \\ 

\noindent\textbf{Stability:} We check that $\nu_R$ is stable over most of our parameter space for the following decay modes  $\nu_R\rightarrow 3 \nu_L$ and $\nu_R\rightarrow \gamma+ \nu_L$, whose widths read \cite{Lee:1977tib,Pal:1981rm,Shrock:1982sc}
\begin{align}
    \Gamma\left(\nu_R\rightarrow 3 \nu_L\right)&= \frac{\sin{(\theta_{LR})}^2}{96\pi^3} \frac{m_{\nu_R}^5}{v_L^4}\,,\nonumber\\
    \Gamma\left(\nu_R\rightarrow \gamma + \nu_L\right)&= \frac{9\alpha_\text{EM} \sin{(\theta_{LR})}^2}{256\pi^4} \frac{m_{\nu_R}^5}{v_L^4}\,,
\end{align}
where $\alpha_\text{EM}\simeq 1/137$ is the electromagnetic fine structure constant. For $m_{\nu_R}>2 m_e$ there also exists the decay $\nu_R \rightarrow \nu_L e^+ e^-$, but here we focus on the sub-MeV regime. We concentrate on the mixing between $\nu_R$ and the lightest $\nu_L$ given by Eq.~\eqref{eq:mix}, because we assume the coupling to the heavier two $\nu_L^{(2,3)}$ is suppressed. 
For $m_{\nu_R}<\mathcal{O}(\SI{10}{\kilo\electronvolt})$, we find that the $\nu_R$ lifetime is larger than the age of the universe. Current \cite{Ng:2019gch,Roach:2019ctw} and next generation \cite{Neronov:2015kca,Dekker:2021bos} X-Ray data is only sensitive to   $m_{\nu_R}>\mathcal{O}(\SI{1}{\kilo\electronvolt})$.

\begin{figure}[t!]
    \centering
    \includegraphics[scale=0.5]{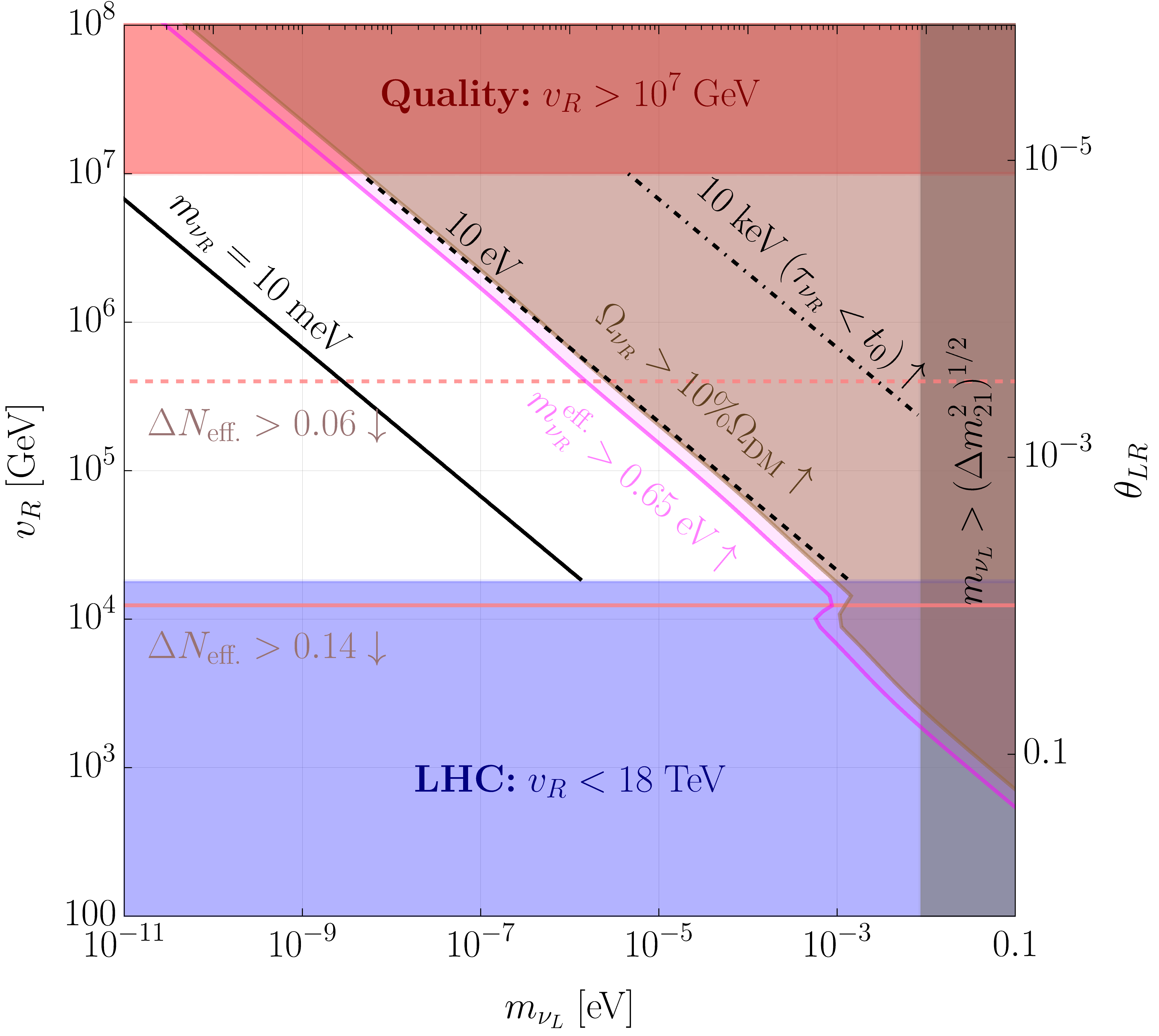}
    \caption{Cosmologically allowed parameter space for the  lightest generation of  $\nu_R$ in the plane spanned by the lightest active neutrino mass $m_{\nu_L}$ and the VEV $v_R$.
    In the gray region our assumption of an effectively massless lightest active neutrino (see the discussion around Eq.~\eqref{eq:light}) breaks down. The red and blue regions are excluded from the limits on $v_R$ from the quality problem (see Eq.~\eqref{eq:qual2}) and collider searches (see Eq.~\eqref{eq:18}). In the brown region, the condition on the relic abundance in Eq.~\eqref{eq:frac} is violated. Above the magenta line, the CMB limit on the effective $\nu_R$ mass in Eq.~\eqref{eq:meff} is violated, which coincides with the aforementioned limit on the relic abundance. We depict iso-contours for $m_{\nu_R}=\SI{10}{\kilo\electronvolt},\; \SI{10}{\electronvolt},\; \SI{10}{\milli\electronvolt}$ and for $\Delta N_\text{eff.}=0.06,\;0.14.$ }
    \label{fig:nuR}
\end{figure}

We depict the parameter space allowed by all precious considerations in Fig. \ref{fig:nuR}. One can see that we require $m_{\nu_R}<\SI{10}{\electronvolt}$ to avoid a too large relic fraction of $\nu_R$. The allowed range of left-right breaking scales  $\SI{18}{\tera\electronvolt}<v_R<\SI{e7}{\giga\electronvolt}$ corresponds to an upper limit on the  lightest active neutrino mass of $m_{\nu_L}< \mathcal{O}((10^{-9}-10^{-3})\;\text{eV})$ and a mixing angle $\mathcal{O}(10^{-5})>\theta_{LR}  > \mathcal{O}(10^{-2})$.\\

\noindent\textbf{Implications for heavier generations:} We can estimate the energy density of the other two heavier $\nu_{R_{2,3}}$ by using the relation in Eq.~\eqref{eq:ratio}. We further assume the lightest active neutrino to be essentially massless and somewhat decoupled from the heavier generation, because we worked in a simplified one-flavor approximation in Section \ref{sec:numass}. This leads to 
\begin{align}\label{eq:sum}
    \sum_{i=2,3} m_{\nu_{R_i}} =  \frac{4}{3}\left(\frac{v_R}{v_L}\right)^2 \sum_{i=2,3} m_{\nu_i}\,.
\end{align}
The sum of active neutrino masses  $\sum_{i} m_{\nu_i}$  is constrained by cosmological data, the Planck collaboration obtained the limit $\sum_i m_{\nu_i} < \SI{0.12}{\electronvolt}$ \cite{Planck:2018vyg}. Combining this with DESI BAO data, the bound tightens to  $\sum_i m_{\nu_i}<\SI{0.072}{\electronvolt}$  \cite{DESI:2024mwx} for a $\Lambda$CDM background. However, one should be careful with this limit, since assuming a background with time dependent dark energy results in a much weaker limit of $\sum_i m_{\nu_i}=0.19^{+0.15}_{-0.18}\;\text{eV}$ \cite{RoyChoudhury:2024wri,RoyChoudhury:2025dhe} (see also Ref.~\cite{Giare:2025ath} for a different view-point on the cosmological neutrino mass bound). We thus conservatively adopt the Planck limit. 

If the heavier two $\nu_R$ states were stable on cosmological timescales, their relic density can at most account for 0.17\% of the observed relic abundance when taking the limit in Eq.~\eqref{eq:18}. In the previous estimate, we assumed that  $\nu_{R_{2,3}}$ are stable on cosmological time-scales. A precise determination of their lifetime relies on the mixing with the active neutrinos, which is beyond our single generation estimate. The decays of $\nu_{R_{2,3}}$ could also be a virtue, as they could be long-lived enough to source an intermediate epoch of early matter domination and the subsequent entropy release from their decays could dilute the energy density of the lightest $\nu_R$ enough so that it becomes compatible with a large enough $m_{\nu_R}$ to comply with the aforementioned bounds  \cite{Bezrukov:2009th}. This approach was used to 
explain the dark matter relic abundance from a diluted population of keV-MeV scale $\nu_R$ in Ref.~\cite{Dror:2020jzy}.

\section{Conclusion}\label{sec:concl}
We conclude by giving an overview of the salient findings of our analysis
\begin{itemize}
    \item In the context of Universal Seesaw model, we show that leptogenesis can work with a single generation of Majorana mediators $N_{L,R}$, as the parity doubling leading to the presence of both $N_L$ and $N_R$ can account for the $CP$ violation from tree-loop-interference. The generalized parity $\mathcal{P}$ needs to be softly broken to allow for different masses of $N_{L,R}$ or else the $CP$ violation vanishes. As a proof of principle, we focused on the coupling to the lightest and effectively massless active neutrino $\nu_L$ and ensured that the heavier generations are sufficiently decoupled. In contrast to previous implementations, our parameter space is barely affected by washout, left-right equilibration and the additional gauge interactions. 

    \item Leptogenesis with high scale left-right symmetry breaking $v_R>M$ needs $M_N>\mathcal{O}(\SI{e10}{\giga\electronvolt})$, but this regime is disfavored by the washout from left-right equilibration in Eq.~\eqref{eq:weak}. Moreover, it can be in tension with the  quality problem of the solution to the strong $CP$ problem. For the case of gauged $\mathcal{P}$,  the dangerous higher dimensional  operators in Eq.~\eqref{eq:qual2} would need to have couplings below $\mathcal{O}(10^{-3})$ to accommodate this range of $v_R$. If $\mathcal{P}$ is instead a global symmetry and the operators in Eq.~\eqref{eq:danger} are realized, one would have to suppress their couplings by at least six orders of magnitude. Consequently, this scenario needs a dedicated  mechanism to sufficiently  forbid these higher dimensional operators for both realizations of generalized parity.
    
    \item Leptogenesis with  $M>v_R$ can work if we explicitly break $\mathcal{P}$ in the neutrino Yukawa couplings to avoid the cancellation of the asymmetries stored in the different lepton chiralities due to the presence of two distinct sphalerons. We find an increase in the mass scale for leptogenesis of $M_N>\mathcal{O}(\SI{e12}{\giga\electronvolt})$, due to the assumption that this breaking
    is small, which further suppresses the amount of $CP$ violation per $N_{L,R}$ decay.
    This hard breaking could arise as a consequence of the higher dimensional operator in Eq.~\eqref{eq:SSB2} together with spontaneous breaking from a parity-odd singlet scalar $S$. 
    However, due to the unavoidable domain walls in the field $S$ for $T_\text{RH}>M_N>v_R$ we can not use a gauged $\mathcal{P}$ for the intended cosmological history, implying that here generalized parity has to be a global symmetry. This region of parameter space is compatible with values of $v_R$ as low as the current limit of 18 TeV \cite{Craig:2020bnv} from collider searches. Thus, our approach is complimentary to the high scale scenarios of Refs.~\cite{Gu:2010yf,Dunsky:2020dhn,Babu:2024glr,Carrasco-Martinez:2023nit}.

    \item Together with the lightest active neutrino, $\nu_L$ we predict a RHN with mass $m_{\nu_R}<\SI{10}{\electronvolt}$ that is produced via relativistic freeze-out from gauge interactions. We find the lightest active neutrino mass in the range,  $m_{\nu_L}< \mathcal{O}((10^{-9}-10^{-3})\;\text{eV})$ together with a active-sterile mixing angle of $\mathcal{O}(10^{-5})>\theta_{LR}  > \mathcal{O}(10^{-2})$. 
    This additional light fermion has an abundance that is small enough not to upset the standard cosmological picture. 
    Using the improved bound of $\Delta N_\text{eff.}\simeq 0.06$ for its contribution to dark radiation, will lead to a lower bound on $v_R$ of $\SI{400}{\tera\electronvolt}$.

    \item The Universal Seesaw without any singlet fermions $N_{L,R}$ can be embedded in the product group $\text{SU}(5)_L\otimes \text{SU}(5)_R$ \cite{Babu:2023dzz}. The presence of $N_{L,R}$ in our model suggests a possible embedding in  $\text{SO}(10)_L\otimes \text{SO}(10)_R$ \cite{Davidson:1987mh,Cho:1993jb}. A potential complication comes from the fact that we take $\mathcal{P}$ to hold to a very good approximation in the quark sector, whereas we have to break it both at the level of the mass spectrum and in the Yukawa couplings of the neutral lepton sector. If generalized parity is broken spontaneously akin to Eqs.~\eqref{eq:SSB1} and \eqref{eq:SSB2}, one then requires a mechanism that generates different Yukawa couplings to the parity breaking field for quarks and leptons despite both sitting in the same multiplet. 
 \end{itemize}

\section*{Acknowledgments}
We are thankful to Yue Zhang for his comments and feedback on the manuscript. KSB is supported in part by the U.S. Department of Energy under grant number DE-SC0016013. MB is supported by \enquote{Consolidación Investigadora Grant CNS2022-135592}, and funded also by \enquote{European Union NextGenerationEU/PRTR}, as well as the Generalitat Valenciana APOSTD/2025 Grant No. CIAPOS/2024/148.  MB would like to thank the CERN theory group, the Lawrence Berkeley National Laboratory as well as the Leinweber Center for Theoretical Physics at the University of California  Berkeley, where parts of this work were completed, for their kind hospitality. SG acknowledges the J.C.~Bose Fellowship (JCB/2020/000011) of the Anusandhan National Research Foundation, and the Department of Space, Government of India. She also acknowledges Oklahoma State University where the work was initiated, for hospitality and Fulbright-Nehru Academic and Professional Excellence fellowship for funding her visit to the US. DV is supported by a McDonald Institute Theory Fellowship funded from the Canada First Research Excellence Fund through the Arthur B. McDonald Canadian Astroparticle Physics Research Institute, and a Subatomic Physics Discovery Grant (individual) from the Natural Sciences and Engineering Research Council of Canada. 

\appendix

\section{Decay Widths and Branching Ratios}\label{app:BRs}
In this appendix we collect all the relevant decay widths and branching fractions. 

\subsection{Generalized Parity Conserving Case}
The decay widths read 
\begin{align}
\Gamma(N_{L} \rightarrow l_L H_L, l_L^\dagger H_L^\dagger) &= \frac{|Y_{LL}|^2}{8\pi} M_{LL}\,,\nonumber\\
\Gamma(N_{L} \rightarrow l_R H_R, l_R^\dagger H_R^\dagger) &= \frac{|Y_{LR}|^2}{8\pi} \left(1-\delta_L\right)^2 M_{LL}\,,\nonumber\\
\Gamma(N_{R} \rightarrow l_L H_L, l_L^\dagger H_L^\dagger) &= \frac{|Y_{LR}|^2}{8\pi} M_{RR}\,,\nonumber\\
\Gamma(N_{R} \rightarrow l_R H_R, l_R^\dagger H_R^\dagger) &= \frac{|Y_{LL}|^2}{8\pi} \left(1-\delta_R\right)^2 M_{RR}\,,
\end{align} 

and the branching fractions are

\begin{align}
    {\rm{Br}}_{L}^{L} &= \frac{\Gamma(N_{L} \rightarrow l_L H_L, l_L^\dagger H_L^\dagger)}{\Gamma(N_{L} \rightarrow l_L H_L, l_L^\dagger H_L^\dagger)+\Gamma(N_{L} \rightarrow l_R H_R, l_R^\dagger H_R^\dagger)}=\frac{1}{2+ (\delta_L-2)\delta_L}\,,\nonumber\\    
    {\rm{Br}}_{L}^{R} &= \frac{\Gamma(N_{L} \rightarrow l_R H_R, l_R^\dagger H_R^\dagger)}{\Gamma(N_{L} \rightarrow l_L H_L, l_L^\dagger H_L^\dagger)+\Gamma(N_{L} \rightarrow l_R H_R, l_R^\dagger H_R^\dagger)}=1-\frac{1}{2+ (\delta_L-2)\delta_L}\,,\nonumber\\ 
    {\rm{Br}}_{R}^{L} &= \frac{\Gamma(N_{R} \rightarrow l_L H_L, l_L^\dagger H_L^\dagger)}{\Gamma(N_{R} \rightarrow l_L H_L, l_L^\dagger H_L^\dagger)+\Gamma(N_{R} \rightarrow l_R H_R, l_R^\dagger H_R^\dagger)} =\frac{r^2}{2 r( r-\delta_L) +\delta_L^2}\,,\nonumber\\
    {\rm{Br}}_{R}^{R} &= \frac{\Gamma(N_{R} \rightarrow l_R H_R, l_R^\dagger H_R^\dagger)}{\Gamma(N_{R} \rightarrow l_L H_L, l_L^\dagger H_L^\dagger)+\Gamma(N_{R} \rightarrow l_R H_R, l_R^\dagger H_R^\dagger)} =1-\frac{r^2}{2 r( r-\delta_L) +\delta_L^2}\,.
\end{align}

\subsection{Generalized Parity Violating Case}
The decay widths read 
\begin{align}
\Gamma(N_{L} \rightarrow l_L H_L, l_L^\dagger H_L^\dagger) &= \frac{|Z_{LL}|^2}{8\pi} M_{LL}\,,\nonumber\\
\Gamma(N_{L} \rightarrow l_R H_R, l_R^\dagger H_R^\dagger) &= \frac{|Y_{LR}|^2}{8\pi}  M_{LL}\,,\nonumber\\
\Gamma(N_{R} \rightarrow l_L H_L, l_L^\dagger H_L^\dagger) &= \frac{|Y_{LR}|^2}{8\pi} M_{RR}\,,\nonumber\\
\Gamma(N_{R} \rightarrow l_R H_R, l_R^\dagger H_R^\dagger) &= \frac{|Z_{RR}|^2}{8\pi}   M_{RR}\,,
\end{align} 
and the branching fractions are
\begin{align}
    {\rm{Br}}_{L}^{L} &=  1-\frac{1}{2+\omega^2+2 \omega \sin{\left(\frac{\varphi-\chi}{2}\right)}}\,,\nonumber\\    
    {\rm{Br}}_{L}^{R} &=  \frac{1}{2+\omega^2+2 \omega \sin{\left(\frac{\varphi-\chi}{2}\right)}}\,,\nonumber\\ 
    {\rm{Br}}_{R}^{L} &=  \frac{1}{2+\omega^2-2 \omega \sin{\left(\frac{\varphi-\chi}{2}\right)}}\,,\nonumber\\
    {\rm{Br}}_{R}^{R} &=  1-\frac{1}{2+\omega^2-2 \omega \sin{\left(\frac{\varphi-\chi}{2}\right)}}\,.
\end{align}

\section{Chemical Potentials}\label{app:chem}
Here we provide the details of our chemical potential analysis for the spectator processes.

\subsection{Below the LR Breaking Scale}\label{app:belowLR}
To compute the sphaleron redistribution coefficient, we follow the treatment of spectator processes of Refs.~\cite{Harvey:1990qw,Dreiner:1992vm,Khlebnikov:1996vj}:
We have to include the $\text{SU}(2)_L$ sphaleron transitions
\begin{align}
    \mu_{l_L} + 3 \mu_{q_L} =0\,.
\end{align}
and ensure that the plasma is hypercharge neutral
\begin{align}
3\left(2\mu_{q_L}-2 \mu_{l_L}  -2 \mu_{u_R}+\mu_{d_R} + \mu_{e_R} \right) + 4  \mu_{H_L} =0\,.
\end{align}
We take all Yukawa interactions to be in thermal equilibrium
\begin{align}
\mu_{q_L}+\mu_{H_L}+\mu_{u_R}&=0\,,\nonumber\\ 
\mu_{q_L}-\mu_{H_L}+\mu_{d_R}&=0\,,\nonumber\\
\mu_{l_L}-\mu_{H_L}+\mu_{e_R}&=0\,,
\end{align}
and for the three fermion flavors of the Standard Model this is valid below $\SI{e5}{\giga\electronvolt}$ \cite{Bodeker:2019ajh}.

In the above, we took generation-independent chemical potentials for the sake of simplicity. For the impact of the different equilibration temperatures for each Yukawa interaction, see Refs.~\cite{Nardi:2005hs,Nardi:2006fx}.
The condition for the QCD sphaleron is already included via the equilibrated quark Yukawa interactions. 

We define the chemical potential baryon and lepton number via 
\begin{align}
    \mu_B &= 2 \mu_{q_L} + \mu_{u_R} + \mu_{d_R}\,,\nonumber\\
    \mu_L &= 2\mu_{l_L} + \mu_{e_R}\,.
\end{align}
and obtain 
\begin{align}
    c_\text{sph.}(T\ll v_R) = \frac{\mu_{B}}{\mu_{B-L}}= \frac{28}{79}\,.
\end{align}

\subsection{Above the LR Breaking Scale}\label{app:aboveLR}

To compute this number, we modified the treatment of Refs.~\cite{Harvey:1990qw,Dreiner:1992vm,Khlebnikov:1996vj}  as follows: 
We include the effect of $\text{SU}(2)_R$ sphalerons in our analysis. For $\text{SU}(2)_L$ the sphaleron rate reads \cite{DOnofrio:2014rug}
\begin{align}
    \Gamma_\text{sph.L}= (18\pm3) \alpha_L^5 T\,,
\end{align}
where 
\begin{align}
    \alpha_L \equiv \frac{g_L^2}{4\pi}\,.
\end{align}
At high scales we expect the gauge couplings of both $\text{SU}(2)$ interactions to be equal ($g_L=g_R$) due to the discrete exchange symmetry  (their renormalization group running is however different, so one expects them to have different values at low scales) and we find that both sphalerons are in thermal equilibrium below 
\begin{align}
    T_\text{sph.L,R} = (0.9-1.2)\times 10^{11}\;\text{GeV} \left(\frac{g_{L,R}}{0.55}\right)^{10}\,,
\end{align}
and consequently we impose
\begin{align}
    \mu_{l_L} + 3 \mu_{q_L} =0, \quad \mu_{l_R} + 3 \mu_{q_R} =0\,.
\end{align}
One the temperature decreases below approximately $v_{L,R}$, the group $\text{SU}(2)_{L,R}$ gets higgsed and its sphalerons become exponentially suppressed  \cite{Kuzmin:1985mm}. 
Applications of the $\text{SU}(2)_R$ sphalerons to baryogenesis can be found in Refs.~\cite{Maleknejad:2020yys,Harigaya:2021txz,Harigaya:2022wzt}.

Note that for scenarios with a conserved lepton number $\mu_{l_L}+\mu_{l_R}=0$, such as Dirac leptogenesis \cite{Dick:1999je} (see also Ref.~\cite{Babu:2024glr} for a recent example based on an extension of the Universal Seesaw), having both sphalerons in thermal equilibrium would lead to a vanishing total asymmetry. 

Additionally, we have to demand the conservation of the total $X$ gauge charge of the plasma (this is the analog of hypercharge conservation in the SM):
\begin{align}
&3\left(\mu_{q_L}- \mu_{l_L} + 2\mu_{U_L}-\mu_{D_L} -\mu_{E_L} -2\mu_{q_R}+ 2 \mu_{l_R} -2 \mu_{U_R}+\mu_{D_R} + \mu_{E_R} \right)\nonumber\\ + &2\left( \mu_{H_L} -\mu_{H_R} \right)=0\,,
\end{align}
where the factor of 3 accounts for the number of fermionic generations.

Furthermore, we take the Yukawa couplings of the charged leptons and quarks to their vector-like partners to be $\mathcal{O}(1)$, so we expect them all to be in equilibrium at high temperatures. This implies for each generation that
\begin{align}
\mu_{q_L}+\mu_{H_L}+\mu_{U_R}&=0,\quad \mu_{q_R} +\mu_{H_R}+\mu_{U_L}=0,\quad \mu_{U_R}+\mu_{U_L}=0\,,\nonumber\\ 
\mu_{q_L}-\mu_{H_L}+\mu_{D_R}&=0,\quad \mu_{q_R} -\mu_{H_R}+\mu_{D_L}=0,\quad \mu_{D_R}+\mu_{D_L}=0\,,\nonumber\\
\mu_{l_L}-\mu_{H_L}+\mu_{E_R}&=0,\quad \mu_{l_R} -\mu_{H_R}+\mu_{E_L}=0,\quad \mu_{E_R}+\mu_{E_L}=0\,.
\end{align}

Once the temperature drops below the masses of the vector-like fermions the Yukawa interactions are given by dimension five operators, that reduce the Standard Model Yukawa couplings once $v_R$ condenses. However, since we generically expect that $M_{LL,RR}$ are much heavier than the vector-like fermions, we can neglect this effect, which would generate a temperature dependence for $c_\text{sph.}(T)$ and we do not expect this to affect our results beyond corrections of at most $\mathcal{O}(10\%)$.

We define the baryon and lepton number chemical potentials as follows 
\begin{align}
    \mu_{L_L}&=2\mu_{l_L}+\mu_{E_L} ,\quad
    \mu_{B_L}=2\mu_{q_L} + \mu_{U_L}+\mu_{D_L}\,,\nonumber\\
    \mu_{L_R}& =2\mu_{l_R}+\mu_{E_R} ,\quad 
    \mu_{B_R}=2\mu_{q_R} + \mu_{U_R}+\mu_{D_R}\,,
\end{align}
and obtain 
\begin{align}
    c_\text{sph.}(T\gg v_R) = \frac{\mu_{B_L}}{\mu_{(B-L)_L}}=\frac{\mu_{B_R}}{\mu_{(B-L)_R}}= \frac{28}{79}\,.
\end{align}

\section{ A New Channel for Depletion of $\Delta_{R}$}\label{sec:WObreak}

Here, we discuss the implications for depletion of $\Delta_R$ in the presence of new fields, for example, a gauge singlet $\psi$, which does not transform under the generalized parity, and just couples to the right-handed lepton and Higgs doublet 
\begin{align}\label{eq:new}
    \mathcal{L}_\psi = m_\psi \psi \psi + Y_\psi l_R H_R \psi + \text{H.c.}\,.
\end{align}
This coupling would then selectively deplete $\Delta_{l_R}$ via scattering with a rate at $T>m_\psi$ given by
\begin{align}
    \Gamma(\nu_R \nu_R \leftrightarrow H_R H_R) \simeq Y_\psi^4 \frac{m_\psi^2}{T}\,,
\end{align}
and via inverse decays 
\begin{align}
    \Gamma(l_R H_R \leftrightarrow \psi) \simeq \frac{Y_\psi^2}{8\pi} \frac{m_\psi^2}{T}\,,
\end{align}
where we assumed $T>m_\psi$, so that the inverse decay is suppressed by the length contraction factor of $m_\psi/T$. 

In order for the selective washout to be efficient we demand that both rates are fast at $T=M_N > m_\psi$, which leads to the following conditions for the scattering process 
\begin{align}
    m_\psi > \SI{2.5e8}{\giga\electronvolt} \left(\frac{0.1}{Y_\psi}\right)^2 \left(\frac{M}{\SI{e10}{\giga\electronvolt}}\right)^\frac{3}{2},
\end{align}
and the inverse decay respectively 
\begin{align}
    m_\psi > \SI{2.5e9}{\giga\electronvolt} \left(\frac{0.1}{Y_\psi}\right) \left(\frac{M}{\SI{e10}{\giga\electronvolt}}\right)^\frac{3}{2}.
\end{align}
Since $\psi$ has large Yukawa couplings to $l_R H_R$, it will rapidly  decay to thermalized bath particles at $T<m_\psi$ and leave no trace behind. 

Note that the new interaction in Eq.~\eqref{eq:new} contributes to the $\nu_R$ mass via a term $v_R^2/m_\psi$. If we consider the following hierarchy: $M_N > m_\psi > v_R$, this new contribution will dominate over the ones in section \ref{sec:numass}.

\bibliographystyle{JHEP}
\bibliography{refs}

\end{document}